\documentclass[a4paper,12pt]{article}
\usepackage[margin=3cm]{geometry}

\usepackage{amssymb}
\usepackage{amstext}
\usepackage{amsmath}
\usepackage{amsthm}

\usepackage[breaklinks]{hyperref}
\usepackage{url}
\usepackage{microtype}
\usepackage{subcaption}
\usepackage{mathtools}
\usepackage{algorithm}
\usepackage{algorithmic}

\usepackage{graphicx}
\usepackage{booktabs}

\usepackage{newtxtext,newtxmath}

\newtheorem{example}{Example}

\usepackage{tikz}
\usetikzlibrary{decorations.text, arrows.meta, positioning}
\usetikzlibrary{decorations.pathreplacing,calc,shapes,positioning}
\usetikzlibrary{shapes,shadows,arrows,positioning,angles,quotes}
\tikzset{My Arrow Style/.style={single arrow, fill=black!15, anchor=base, align=center,text width=2.3cm}}
\tikzstyle{arrow} = [thick,->,>=stealth]
\usetikzlibrary{calc,trees,positioning,arrows,fit,shapes,calc}

\tikzstyle{startstop} = [rectangle, rounded corners, minimum width=1.5cm, minimum height=0.5cm,text centered, draw=black, fill=red!30]
\tikzstyle{io} = [trapezium, trapezium left angle=70, trapezium right angle=110, minimum width=0.5cm, minimum height=0.5cm, text centered, draw=black, fill=blue!30]

\tikzstyle{process} = [rectangle, minimum width=3cm, minimum height=0.5cm, text centered, draw=black, fill=orange!30]
\tikzstyle{decision} = [diamond, minimum width=0.5cm, minimum height=0.1cm, text centered, draw=black, fill=green!30]
\tikzstyle{process2} = [rectangle, minimum width=1cm, minimum height=0.5cm, text centered, draw=black, fill=orange!30]
\tikzstyle{arrow} = [thick,->,>=stealth]

\usetikzlibrary{automata, positioning}
\usetikzlibrary{shapes.geometric, arrows, positioning}

\tikzstyle{block} = [rectangle, draw, text centered, rounded corners, minimum height=2em]
\tikzstyle{line} = [draw, -stealth, thick]
\tikzstyle{cloud} = [ellipse, draw, text centered, minimum height=2em, thick]
\tikzstyle{dashedcloud} = [ellipse, draw, dashed, text centered, minimum height=2em, thick]
\usetikzlibrary{positioning, arrows.meta, shapes.geometric, decorations.pathreplacing}

\theoremstyle{definition}
\newtheorem{definition}{Definition}[section]

\begin{document}

\title{Testing for Coordination Failures in Cooperative Multi-Agent Reinforcement Learning}

\author{Dennis Gross, Jørn Eirik Betten, Helge Spieker*}
\date{Simula Research Laboratory, Oslo, Norway\\Corresponding author: helge@simula.no}

\maketitle

\begin{abstract}
\emph{Deep parallel cooperative multi-agent reinforcement learning (CMARL)} trains multiple agents that all act at every time step to optimize a common objective.
Like single-agent RL, CMARL can produce unsafe behaviors that violate implicit safety requirements not captured by the reward, and these behaviors are hard to explain.
However, testing CMARL policies for safety has received less attention than the safety testing of single-agent RL policies.
In this paper, we propose \emph{DisQ}, a test case selection method that identifies failure-revealing states in value-based CMARL by exploiting a signal specific to trained Q-learning agents.
The signal is the \emph{Q-value gap}, an explainable metric that flags a state when the agents' estimates of the achievable return diverge by more than a calibrated threshold.
DisQ selects such states for execution and skips those without a gap.
Our experiments in four CMARL benchmark environments show that, in most of the evaluated configurations, DisQ improves failure detection over single-agent RL testing methods applied to CMARL, namely random testing and search-based testing with genetic algorithms.
Our main contributions are the Q-value gap, an explainable metric for CMARL miscoordination, and DisQ, a test case selection method that adapts existing single-agent testing methods to CMARL.
\end{abstract}


%\linenumbers

\section{Introduction}
\emph{Deep parallel cooperative multi-agent reinforcement learning (CMARL)}
extends \emph{single-agent reinforcement learning (RL)} to distributed
sequential decision-making problems in which multiple agents act
simultaneously toward a common objective. Such problems arise in domains like
manufacturing job scheduling, autonomous driving, and
gaming~\cite{DBLP:journals/tvt/GanLLT24,DBLP:journals/aei/GuiZTZZ24,DBLP:journals/corr/abs-1912-06721},
where CMARL scales better than single-agent RL, since per-agent policies avoid
the joint action space that grows exponentially in the number of agents~\cite{DBLP:conf/icml/RashidSWFFW18}, and has been reported to reach higher returns in these settings~\cite{DBLP:journals/rcim/ZhangZTZG22,DBLP:journals/corr/abs-1912-06721}.
Agents act in parallel, each selects and executes an action at every time step according to its own learned policy, and all receive a joint reward afterwards~\cite{wong2022deep}, as opposed to turn-based settings in which the agents act one after another.
We further focus on \emph{value-based} CMARL, where each agent's policy is derived from a learned Q-function that estimates the expected cumulative reward of each action in a given state~\cite{Watkins1992,mnih2013playing}.

Like their single-agent counterparts, trained CMARL policies can exhibit
\emph{unsafe behaviors}~\cite{DBLP:conf/aips/GrossS0023}, such as collisions between agents~\cite{DBLP:journals/dcan/LiLLX24}: reward functions rarely capture all safety requirements~\cite{DBLP:journals/aamas/VamplewSKRRRHHM22}, and the neural networks representing the policies hide the details that drive their decisions~\cite{DBLP:journals/ml/Bekkemoen24}.
Testing is the established answer to this problem.
For single-agent RL, a mature line of \emph{search-based software testing (SBST)} methods
exists that searches the space of environment configurations for failure-revealing test
cases~\cite{DBLP:conf/ast/MazouniSGA24,DBLP:conf/issta/PangYW22,DBLP:journals/tosem/BiagiolaT24,DBLP:conf/ecai/ZhouL23,DBLP:conf/issta/EniserGW0C22}, in some cases guided by explainable-RL
techniques~\cite{DBLP:journals/tosem/BiagiolaT24}.

For CMARL, the picture is different in two respects.
First, existing CMARL testing work has largely focused on \emph{adversarial robustness}, i.e., perturbing the agents' observations or actions to degrade performance~\cite{DBLP:conf/ecai/ZhouL23,DBLP:conf/cvpr/GuoCHYYL22,DBLP:conf/issta/MaWWXWLX024}.
This addresses security-related threats, but not failures that originate in the agents' learned coordination itself: states in which the agents, unperturbed, fail to cooperate.
Second, testing is more expensive than in the single-agent case, because the joint state and action spaces grow exponentially with the number of agents~\cite{DBLP:journals/tosem/BiagiolaT24,DBLP:conf/aips/GrossS0023}, so every executed test episode counts.
Applying single-agent SBST methods to CMARL unchanged is possible, but ignores the one signal that is specific to cooperative agents: whether their learned expectations about the joint future agree.

This work exploits exactly that signal.
Since cooperative agents are trained on a shared reward, their Q-functions should assign approximately the same value to the same situation.
We define the \emph{Q-value gap}, an explainable CMARL metric that flags a state when the agents' maximal Q-values diverge by more than a calibrated threshold.
A Q-value gap means the agents hold contradictory expectations about the future, an indicator of potential miscoordination.
We show that gap states are disproportionately failure-revealing under a generic reward-based oracle available in every environment we study, and identify the conditions under which they are not.

Building on this metric, we introduce \emph{Disagreeing Q-values (DisQ)}, a test case selection method for value-based CMARL.
Given a candidate initial state from an arbitrary test case generator, DisQ computes the Q-value gap at that state and selects the test case for execution only if a gap is present.
The decision requires a single forward pass per agent and no interaction with the environment.
This distinguishes DisQ from adversarial CMARL testing, which perturbs the system under test, and from
diversity-driven multi-agent testing such as MASTest~\cite{DBLP:conf/issta/MaWWXWLX024}, which requires executing episodes and analyzing the resulting trajectories before assessing a test case.
To the best of our knowledge, DisQ is the first method to select CMARL test cases from the agents' Q-values at the initial state alone.
Because DisQ only filters candidate states, it is compatible with existing test case generators.
For example, we evaluate it on top of random and genetic SBST in our experiments.

DisQ rests on three assumptions:
(i) agents whose policies are derived from per-agent action-value functions, as in deep Q-learning; pure policy-gradient methods expose no value estimates, and common actor-critic CMARL architectures employ shared or joint-action critics for which inter-agent disagreement is not meaningful;
(ii) access to the agents' Q-values, i.e., a white-box testing setting, as is available to developers testing their own systems before deployment;
and (iii) environments in which the initial configuration (and hence the agents' coordination at the start of an episode) influences the episode outcome, since DisQ evaluates only initial states.
We discuss (i) and (ii) in Section~\ref{sec:limitations}, probe the boundary of (iii) empirically in Section~\ref{sec:limits}, and return to the failure class in Section~\ref{sec:threats}.

Our contributions are:
\begin{itemize}
    \item the \emph{Q-value gap}, an explainable metric for potential miscoordination in value-based CMARL, computable from a single state without executing the system;
    \item \emph{DisQ}, a test case selection method based on the Q-value gap that composes with existing SBST test case generators and avoids executing test cases that are unlikely to fail;
    \item an evaluation in four benchmark environments under both centralized and decentralized training, showing that, in most configurations, DisQ-selected test cases expose more failures and lower rewards than those of vanilla random and genetic SBST, with statistically significant differences.
\end{itemize}

The remainder of the paper is structured as follows.
Section~\ref{sec:background} introduces single- and multi-agent reinforcement learning, Q-learning, and search-based testing, including the definitions of test case and failure used throughout.
Section~\ref{sec:method} presents the Q-value gap and the DisQ selection method, together with their limitations.
Section~\ref{sec:experiments} describes the experimental setup, and Section~\ref{sec:results} presents the results along four research questions.
Section~\ref{sec:threats} discusses threats to validity, Section~\ref{sec:relwork} reviews related work, and Section~\ref{sec:conclusion} concludes the paper.

\section{Background}\label{sec:background}

\subsection{Reinforcement Learning}\label{sec:rl}
A \textit{probability distribution} over a set $X$ is a function $\mu : X \rightarrow [0,1]$ with $\sum_{x \in X} \mu(x) = 1$. The set of all distributions on $X$ is denoted $Distr(X)$.

\begin{definition}[Markov Decision Process]\label{def:mdp}

\noindent A \emph{Markov decision process (MDP)} is a tuple $M = (S,s_0,Act,Tr, rew)$ where
$S$ is a finite, nonempty set of states; $s_0 \in S$ is an initial state; $Act$ is a finite set of actions; $Tr\colon S \times Act \rightarrow Distr(S)$ is a partial probability transition function;
$rew \colon S \times Act \rightarrow \mathbb{R}$ is a reward~function.
\end{definition}
We employ a factored state representation where each state $s$ is a vector of features $(f_1, f_2, ...,f_n)$ where each feature $f_i\in \mathbb{R}$ for $1 \leq i \leq n$ ($n$ is the dimension of the state).
We define $F$ as the set of all features $f_i$ in state $s \in S$.
The available actions in $s \in S$ are $Act(s) = \{a \in Act \mid Tr(s,a) \neq \bot\}$ where $Tr(s, a) = \bot$ denotes that no transition is defined for action $a$ in state $s$, i.e., $a$ is unavailable in $s$.

A \emph{policy} is the rule an agent follows when deciding which action to take, based on its \emph{observation} of the environment state.
Here, the observation might equal the environment state in case of a fully observable MDP, or it might only be a subset of the information, e.g., each agent's individual view or when the environment is only partially observable.

\begin{definition}[Observation]
    We define the observation function $\mathbb{O} \colon S \rightarrow O$ as a function that maps a state $s \in S$ to an observation $o \in O$.
    An observation $o \in O$ is a vector of features $f_j$ from state $s$.
    An observation composed of all features $(f_1, f_2, ...,f_n)$  is a complete observation.
\end{definition}

\begin{definition}[Policy]
A \emph{memoryless deterministic policy} for an MDP $M$ is a function $\pi \colon O \rightarrow Act$ that maps an observation $o \in O$ to an action $a \in Act$.
\end{definition}

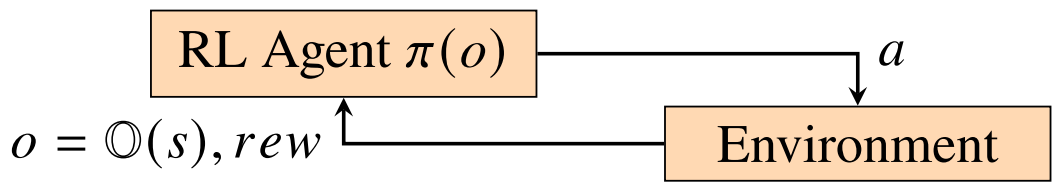
\begin{figure}[t]
\centering
\scalebox{0.85}{
    \begin{tikzpicture}[]
     {};
    \node (agent1) [process] {RL Agent $\pi(o)$};
    \node (env) [process, below of=agent1,yshift=0.3cm,xshift=4cm] {Environment};

    \draw [arrow] (agent1) -| node[anchor=west] {$a$} (env);
    \draw [arrow] (env) -| node[anchor=east] {$o = \mathbb{O}(s),rew$} (agent1);
    \end{tikzpicture}
    }

\caption{Single-agent RL system where an agent receives an observation $o$ and a reward $rew$ from the environment after an action $a$ was applied to the environment.}
\label{fig:rl}
\end{figure}

For MDPs with many states and transitions, obtaining optimal policies $\pi^*$ is difficult.
The standard learning goal for RL is to find a policy $\pi$ in an MDP such that $\pi$ maximizes the expected discounted reward, that is, $\mathbb{E}[\sum^{L}_{t=0}\gamma^t R_t]$, where $\gamma$ with $0 \leq \gamma \leq 1$ is the discount factor, $R_t$ is the reward at time $t$, and $L$ is the total number of steps (see Figure~\ref{fig:rl}).

\subsection{Q-learning}\label{sec:qlearning}
Q-learning~\cite{Watkins1992} learns an action-value function $Q(s,a)$ estimating the expected discounted return of taking action $a$ in state $s$ and acting greedily thereafter.
After observing a transition $(s, a, r, s')$, the estimate is updated toward the temporal-difference (TD) target:
\[
Q(s,a) \leftarrow Q(s,a) + \alpha \big( r + \gamma \max_{a'} Q(s', a') - Q(s,a) \big),
\]
where $\alpha$ is the learning rate and $\gamma$ the discount factor.
In \emph{tabular} Q-learning, $Q$ is stored explicitly per state–action pair, which is infeasible for large or continuous state spaces.
\emph{Deep} Q-learning (DQN)~\cite{mnih2013playing} instead approximates $Q$ with a neural network $Q_\theta$, trained by minimizing the squared TD error over mini-batches of transitions drawn from a replay buffer, typically with a periodically updated target network for stability.
Throughout this paper, Q-values refer to the outputs of such trained networks; consequently, they are \emph{estimates} whose quality depends on training duration and hyperparameters.

\subsection{Cooperative Multi-agent Reinforcement Learning}\label{sec:cmarl}

\begin{figure}[t]
\centering
\scalebox{0.85}{
 \begin{tikzpicture}[]
    \node (agent1) [process] {RL Agent 1 $\pi_1(o_1)$};
    \node (agent2) [process, right of=agent1,xshift=2.75cm] {RL Agent 2 $\pi_2(o_2)$};
    \node (action_product) [io, below of=agent1,xshift=1.75cm,yshift=0cm] {$a=a_1 \times a_2$};
    \node (env) [process, below of=action_product,yshift=-0cm] {Environment};

    \node (obs) [io, below of=env,yshift=0cm] {$s,rew$};
    \node (obs1) [io,left of=obs,xshift=-3cm] {$o_1$, $rew$};
    \node (obs2) [io,left of=obs,xshift=5.2cm] {$o_2$,  $rew$};

    \draw [arrow] (agent1) |- node[anchor=east] {$a_1$} (action_product);
    \draw [arrow] (agent2) |- node[anchor=west] {$a_2$} (action_product);
    \draw [arrow] (action_product) -- node[anchor=west] {$a$} (env);
    \draw [arrow] (env) -- node[anchor=south] {} (obs);
    \draw [arrow] (obs) -- node[anchor=south] {} (obs1);
    \draw [arrow] (obs1) |- node[anchor=south] {} (agent1);
    \draw [arrow] (obs) -- node[anchor=south] {$o_1 = \mathbb{O}_1(s)$} (obs1);
    \draw [arrow] (obs) -- node[anchor=north] {$rew$} (obs1);
    \draw [arrow] (obs) -- node[anchor=south] {} (obs2);
    \draw [arrow] (obs) -- node[anchor=south]  {$o_2 = \mathbb{O}_2(s)$} (obs2);
    \draw [arrow] (obs) -- node[anchor=north]  {$rew$} (obs2);
    \draw [arrow] (obs2) |- node[anchor=south] {} (agent2);
    \end{tikzpicture}
}
    \caption{Cooperative MARL setting with two agents performing a joint action $a=a_1\times a_2$, leading to a new state $s$ and shared reward $rew$, divided into observations $o_1$ and $o_2$.}
    \label{fig:cmarl}

\end{figure}
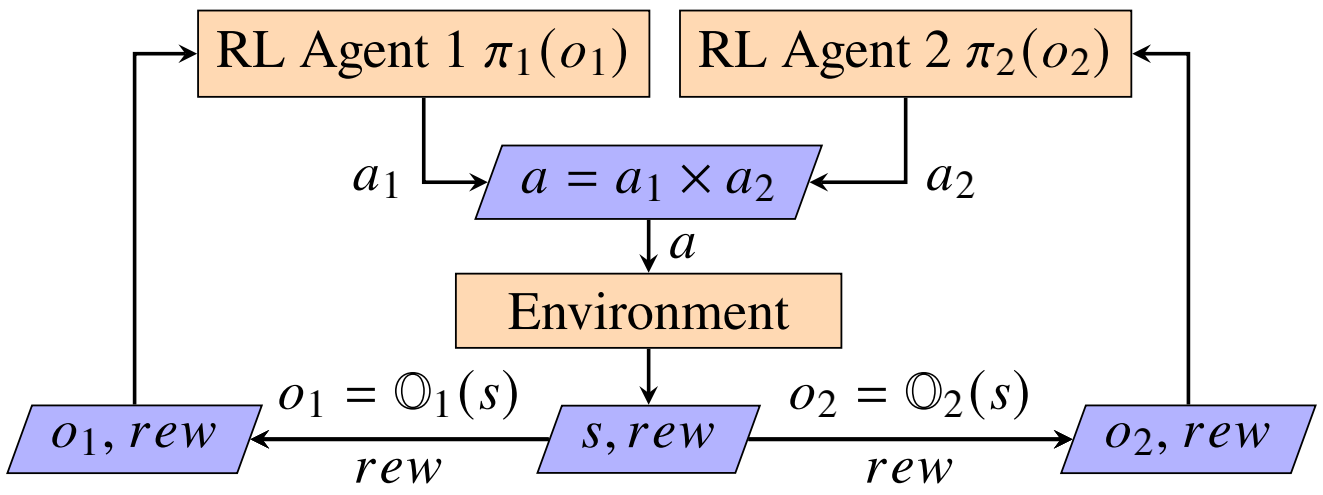

CMARL extends the RL idea to find the near-optimal policies $\pi_i$ in a Cooperative Multi-Agent Markov Decision Process (CMMDP) setting (see Figure~\ref{fig:cmarl} and Definition~\ref{def:cmmdp}).
Each agent policy $\pi_i$ is represented by a neural network.
A neural network is a function parameterized by weights $\theta_i$.
The neural network policy $\pi_i$ can be trained by minimizing a sequence of loss functions $J(\theta_i,o_i,a_i)$~\cite{mnih2013playing}.

In CMARL, local and shared rewards each shape what the agents learn~\cite{DBLP:journals/aamas/ZhuDW24}.
The agents act in lock-step, each taking their individual action, and receive rewards afterwards.
With local rewards, each agent learns based on its own performance, which may not align with the overall cooperative goal. Shared rewards, however, provide all agents with the same feedback, promoting coordination toward a common objective.
This can lead to challenges like credit assignment, where it is hard to pinpoint which agent contributed most to the outcome.
Balancing both reward types allows CMARL to combine individual specialization with collaboration, enabling the discovery of near-optimal policies.

\begin{definition}[CMMDP]\label{def:cmmdp}
    A cooperative multi-agent Markov decision process (CMMDP) is a tuple $\mathcal{M} = (S, s_0, I, Act \coloneqq \{Act_i\}_{i \in I}, T, rew, \{O_i\}_{i \in I}, \{\mathbb{O}_i\}_{i \in I})$ where $S$ is a finite, nonempty set of states; $s_0 \in S$ is an initial state; $I$ is a finite, nonempty set of agents; $Act_i$ is a finite, nonempty set of actions available to agent $i$; $T \colon S \times Act_1 \times ... \times Act_{|I|} \rightarrow Distr(S)$ is a transition function, and $rew \colon S \times Act \rightarrow \mathbb{R}$ is a joint reward function.
    Each agent $i$ gets its local partial observation $o_i \in O_i$ according to the observation function $\mathbb{O}_i(s) \colon S \rightarrow O_i$.
    The observations of all agents result in the full state.
\end{definition}

\subsection{Centralized and Decentralized CMARL Training}
CMARL policies can be trained in two regimes, both of which we test.
In \emph{centralized} training, all agents share the same Q-network, which is trained across all agents simultaneously.
The multi-agent environment is converted into a vectorized environment whose parallel instances correspond to
the individual agents, so one set of weights is updated from every agent's transitions.
During execution, each agent $i$ passes its own observation $o_i$ through these shared weights and selects an action from its own action space $Act_i$; the output head spans $\max_{i \in I} |Act_i|$ actions.
Where the agents' action spaces differ in size, the smaller ones are padded to this maximum, and an index that is out of range for the acting agent is remapped to that agent's default action rather than being masked out before selection.
Among the environments used in the experimental evaluation, this applies only to Simple Speaker Listener.
In all other environments, all agents have action spaces of equal size, so no padding is triggered.
Centralized training has the advantage that it is more data-efficient, but it is not specialized to each agent role in the environment.
In \emph{decentralized} training (independent Q-learning), each agent $i$ trains its own Q-network $Q_i(o_i, a_i)$ on its local observation, with no parameter sharing between agents, treating the other agents as part of the environment.
Decentralized training allows for better specialization on each agent's role, but it scales linearly with the number of agents and usually requires more training iterations to converge to a stable policy.
We evaluate DisQ under both regimes because they produce structurally different Q-functions, and hence potentially different Q-value gap behavior.

\subsection{Search-based Testing Methods}
Testing complex software systems such as CMARL agents is challenging because their behavior in the environment is hard to predict analytically.
\emph{Search-Based Software Testing (SBST)} methods can address these challenges by formulating the testing process as an optimization problem, where the goal is to find input test cases that are most likely to reveal failures in the system~\cite{mcminn_stvr04}.

In the context of CMARL, the \emph{system under test} consists of the trained CMARL agents in the CMMDP $\mathcal{M}[\pi_I]$ with the trained agent policies $\pi$ associated with the agents $I$.

\begin{definition}[Test case]\label{def:testcase}
Given a CMMDP $\mathcal{M}$ and trained policies $\pi_I$, a \emph{test case}
$\tau \in \mathcal{T}$ is an initial state $s_0 \in S_0 \subseteq S$, where
$S_0$ denotes the set of valid initial configurations of the environment
(e.g., agent and landmark positions). \emph{Executing} $\tau$ means running
$\mathcal{M}[\pi_I]$ from $s_0$ for one episode of at most $L$ steps,
yielding the accumulated reward $R(\tau) = \sum_{t=0}^{L} rew(s_t,a_t)$ and a
trace of environment events.
\end{definition}

\begin{definition}[Failure]\label{def:failure}
A test case $\tau$ \emph{fails} if the accumulated reward falls below a reference threshold $\theta$, i.e., $R(\tau) < \theta$, where $\theta$ is derived from the reward distribution of $\pi_I$ over $N$ validation episodes sampled uniformly from $S_0$.
\end{definition}

The threshold $\theta$ can be instantiated in different ways.
A parametric choice is $\theta = \mu - k\sigma$, where $\mu$ and $\sigma$ are the mean and standard deviation of the validation rewards and $k$ is a user-defined severity parameter.
A non-parametric alternative sets $\theta$ to a low quantile of the validation reward distribution, which requires no distributional assumption and makes failure rates comparable across environments with different reward scales.
We use the parametric instantiation ($k = 2$) to characterize the robustness of the policies under test (Section~\ref{sec:experiments}) and the quantile instantiation for the method comparison (Section~\ref{sec:metrics}), because the two settings ask different questions.
Robustness is a statement about a single distribution, and $\theta = \mu - k\sigma$ places the threshold at an interpretable distance from that policy's own mean, so ``fails in $2\%$ of episodes at $k=2$'' has an absolute meaning.
That same property makes it unsuitable for comparing methods: at a fixed $k$ the base rate depends on the shape of each policy's reward distribution, so enrichment measured against it is not comparable across environments.
The quantile instantiation instead pins the baseline failure rate per policy under test, and assumes no particular distributional form.

We define failure through the accumulated reward, rather than through an environment-specific termination condition, because the comparison requires an oracle that is available in every environment under study and graded rather than binary.
In our experiments, \emph{Simple Spread} and \emph{Simple Speaker Listener} define no terminal failure condition at all and always run to a fixed horizon\footnote{In Knight Wizard Zombies and Coin Collector, where terminal conditions do exist, they fired in all episodes we recorded, assigning the same failure rate to all testing methods.}. 
RQ2 relates the magnitude of the Q-value gap to the severity of the exposed behavior, which a binary condition cannot express.

\begin{definition}[Coordination failure]\label{def:coordfailure}
A failure is a \emph{coordination failure} if it arises from incompatible action choices among the agents rather than from an individual agent's inability to act well in isolation, i.e., the joint behavior violates the shared objective, although each agent acts greedily with respect to its own value estimates.
\end{definition}

Definition~\ref{def:coordfailure} names the failure class DisQ targets; Definition~\ref{def:failure} is what our evaluation measures.
The reward-based criterion rests on the accumulated reward alone, so it cannot attribute a failure to incompatible choices between agents rather than to an individual agent acting poorly.
Our empirical claims are scoped accordingly: DisQ exposes lower rewards and more reward-threshold failures.
Section~\ref{sec:threats} quantifies how closely those failures coincide with observable coordination events.
Environment-specific unsafe events, such as collisions between agents, are penalized by the reward functions of our environments and are therefore reflected in Definition~\ref{def:failure} rather than treated as a separate failure condition.
Dedicated failure criteria, e.g., explicit unsafe-event oracles in safety-critical deployments, can be substituted without changing the selection method.

Given these failure definitions, SBST can be formalized as an optimization problem in the form of:
\begin{equation*}
\tau^* = \arg\max_{\tau \in \mathcal{T}} F(\tau, \pi_I)
\end{equation*}
where \( \tau^* \) is the optimal test case, and \( F(\tau, \pi_I) \) is the fitness function that quantitatively measures how close a given test case is to revealing a failure in the CMARL system.
In our context, the fitness is the negated accumulated reward \(-R(\tau)\), so that maximizing $F$ drives the search toward episodes that fall below the failure threshold of Definition~\ref{def:failure}.

SBST requires a search technique.
Popular methods are \emph{genetic algorithms}~\cite{DBLP:journals/tosem/BiagiolaT24} and other metaheuristic optimization methods~\cite{DBLP:conf/ast/MazouniSGA24}, which systematically explore the input space.
A genetic algorithm is an optimization technique inspired by natural selection, where a population of candidate solutions evolves over time.
The algorithm applies genetic operators such as selection, crossover, and mutation to create new generations of solutions to improve their quality according to a defined fitness function~\cite{DBLP:conf/ast/MazouniSGA24}.
In the context of testing, the genetic algorithm modifies the test inputs, i.e., environment parameters such as the agents' starting positions, to minimize the expected performance according to the failure criterion.

\section{Methodology}\label{sec:method}
This section introduces the test case selection method called \emph{Disagreeing Q-values (DisQ)}.
This method selects test cases by analyzing the difference between the agents' maximal Q-values at given initial states without requiring additional information or execution of the actual episodes.
Because the agents are trained on a shared reward, their maximal Q-values should roughly agree. Large discrepancies, i.e., Q-value gaps, therefore pinpoint test cases where coordination is likely to fail (Example~\ref{ex:disq}).
DisQ is independent of the source of the test case states and is compatible with many test case generators, for example, random sampling of initial states or genetic algorithms.

DisQ evaluates the Q-value gap only at the initial state. This choice is deliberate.
Selection becomes execution-free, i.e., it requires a single forward pass per agent, but no involvement of the environment, such that non-promising test cases can be discarded without any simulation cost, which is precisely where CMARL testing is expensive.
The trade-off is that DisQ cannot detect miscoordination that emerges only later in an episode: agents whose value estimates agree at $s_0$ may still diverge at step $t > 0$.
Extending DisQ to monitor $D(s_t)$ during execution, e.g., as an online abort/prioritization criterion, is a natural extension, but it forfeits the execution-free property that motivates the method.
We quantify how much signal this trade-off forgoes, and for which training regimes online monitoring would be selective, in Section~\ref{sec:midepisode}.

\begin{example}\label{ex:disq}
    To illustrate how DisQ works, consider a simple scenario with two CMARL agent policies.
    One policy estimates that choosing action 1 from a particular state will maximize the expected cumulative reward to $100$. In contrast, another policy in the same state predicts a maximum expected cumulative reward of only $20$ for its best action.
    This large divergence between the agents' most optimistic outcomes is called the Q-value gap.
    It indicates a contradiction, as both agents were trained on a common reward function and should predict approximately the same cumulative reward from a particular state.
\end{example}
The following subsections introduce the concept of Q-value gap, then discuss how to apply it as a test case selection method to SBST methods, and discuss its limitations.

\subsection{Q-value Gap}
This metric provides explainability for CMARL by identifying differences in agents' maximal Q-values.
Measuring the total distance between agents' maximal Q-values reveals states where the agents may miscoordinate and fail to generalize their learned coordination policies to new or less common states.

We consider a system of $|I|$ parallel cooperating agents with shared reward, where each agent $i \in \{1, 2, \dots, |I|\}$ has an associated Q-function denoted by $Q_i: O_i \times Act_i \rightarrow \mathbb{R}$.

For a given state $s \in S$, we define the maximal Q-value for the agent $i$ as:
$$
Q_i^{\max}(\mathbb{O}_i(s)) = \max_{a \in Act_i} Q_i(\mathbb{O}_i(s), a)
$$

We define the \emph{total distance} $D(s)$ among all agents at state $s$ as the sum of pairwise distances, i.e., absolute differences, between the maximal Q-values of all distinct agent pairs:
\[
D(s) = \sum_{i=1}^{|I|} \sum_{\substack{j=1 \\ j > i}}^{|I|} d(Q_i^{\max}(\mathbb{O}_i(s)), Q_j^{\max}(\mathbb{O}_j(s)))
\]

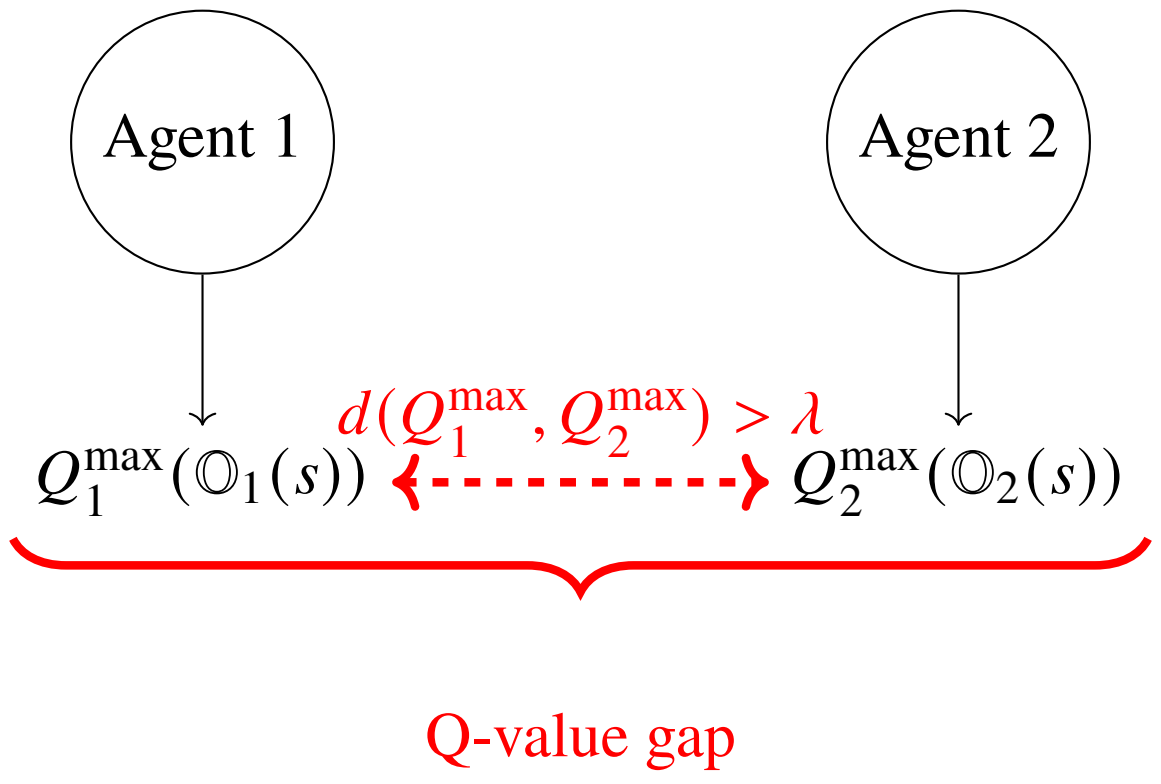
\begin{figure}[t]
\centering
\begin{tikzpicture}[node distance=5cm, auto]
    \node[circle, draw, minimum size=1cm] (A1) {$\text{Agent }1$};
    \node[circle, draw, minimum size=1cm, right of=A1] (A2) {$\text{Agent }2$};

    \node[below=1cm of A1] (Q1) {$Q_1^{\max}(\mathbb{O}_1(s))$};
    \node[below=1cm of A2] (Q2) {$Q_2^{\max}(\mathbb{O}_2(s))$};

    \draw[->] (A1) -- (Q1);
    \draw[->] (A2) -- (Q2);

    \draw[<->, dashed, red, ultra thick] (Q1) -- node[above] {$d(Q_1^{\max}, Q_2^{\max}) > \lambda$} (Q2);

    \node[below=1cm of Q1, xshift=2.5cm, text=red] (CF) {Q-value gap};

    \draw [decorate,decoration={brace, amplitude=10pt, mirror}, red, ultra thick]
    (Q1.south west) -- (Q2.south east) node[midway, below=15pt] {};

\end{tikzpicture}
\caption{Illustration of Q-value gap between two agents when the distance between their maximal Q-values exceeds the threshold \(\lambda\).}
\label{fig:coordination-failure-two-agents}
\end{figure}

\begin{definition}[Q-value gap]
    A \emph{Q-value gap} occurs at a state $s \in S$ if the total distance $D(s)$ exceeds a predefined threshold $\lambda > 0$. Formally,
    \[
    D(s) > \lambda
    \]
    where $\lambda$ is a positive constant determining the sensitivity of detecting Q-value gaps (see also Figure~\ref{fig:coordination-failure-two-agents}).
\end{definition}
In this work, the distance always refers to the absolute difference between two Q-values, which are represented as floating point numbers.

\subsection{Disagreeing Q-values (DisQ)}\label{sec:DisQ}

\begin{algorithm}[t]
\caption{DisQ Test Case Selection Algorithm\label{alg:disq}}
\begin{algorithmic}[1]
\STATE \textbf{Inputs:} Distance threshold $\lambda$, Environment CMMDP, Test case generator TCG
\WHILE{test budget not exceeded}
    \STATE Generate new initial state $s$ for MDP by user-defined TCG
    \STATE Measure Q-values for all agents
    \STATE Calculate $D(s)$ for Q-values
    \IF{$D(s) > \lambda$}
        \STATE Select test case for execution
        \STATE (Optional) Collect metrics for test case generator
    \ENDIF
\ENDWHILE
\end{algorithmic}
\end{algorithm}
DisQ applies the Q-value gap as a selection criterion: it flags test cases likely to fail before executing them.
DisQ focuses on states where Q-value gaps arise, as these can often reveal problematic scenarios within the environment (see Section~\ref{sec:worse}).
A simple version of how DisQ can be applied is outlined in Algorithm~\ref{alg:disq}.
Lines 4–6 are the core of DisQ; the rest depends on the testing setup.

Before testing, we calibrate $\lambda$ by executing the CMARL policies $n$ times in the environment, measuring the total $D(s)$ at each initial state, and then setting $\lambda$ to the $p$-th quantile of the observed total $D(s)$ values.
The quantile $p$ trades selectivity against throughput.
A higher quantile selects fewer, more suspicious test cases, but as $p \to 100$ the selection becomes so restrictive that the generator discards nearly all candidates, and the effective number of executed tests within a fixed budget collapses.
The number of calibration episodes $n$ in turn determines the stability of the quantile estimate: too few episodes yield a noisy $\lambda$ and thus an unreliable selection boundary, while every calibration episode is subtracted from the test budget available for actual testing.
Both parameters are therefore budget-dependent choices rather than properties of the method.

After the calibration, and at the beginning of each test case, DisQ evaluates whether a Q-value gap occurs at the initial state.
When a Q-value gap is detected ($D(s) > \lambda$), the test executes the agents' policies in the environment to further investigate the system's behavior to potentially detect a failure per Definition~\ref{def:failure}.
If no Q-value gap is found, the generator produces a new initial state and repeats the test case selection process.

DisQ can be applied as a test case selection method to SBST methods such as random search or genetic algorithm.
In the random search SBST method, we apply the test case selection at the initial state without executing the full test case.
Depending on the Q-value gap, an execution of the test case could then be skipped to avoid the computational cost of a non-promising test case execution.
For SBST with genetic algorithms, we give individuals the highest possible (worst) fitness when the individual's initial state is not a Q-value gap state.

\subsection{Limitations}\label{sec:limitations}
DisQ's applicability is bounded by three assumptions:
it requires per-agent action-value estimates (i),
access to those estimates (ii),
and environments in which initial-state coordination influences the episode outcome (iii).

\paragraph{Per-agent Action-value Estimates.}
The Q-value gap compares the agents' \emph{independent} expectations of the achievable return, so DisQ requires each agent to expose an action-value estimate over its own action space.
Deep Q-learning provides this natively.
Pure policy-gradient methods such as PPO~\cite{DBLP:journals/corr/SchulmanWDRK17} do not: they output
probability distributions over actions rather than return estimates, and two agents may assign similar probabilities to their preferred actions while expecting very different returns, so a gap over probabilities would not carry the same meaning.
Actor-critic CMARL methods do train value functions, but their mainstream architectures do not expose the quantity DisQ needs.
MAPPO~\cite{DBLP:conf/nips/YuVVGWBW22} learns a centralized \emph{state-value} critic $V(s)$, which provides no per-action values; moreover, in the standard configuration the critic parameters are shared across agents, so any
disagreement between the agents' value estimates is zero by construction.
COMA~\cite{DBLP:conf/aaai/FoersterFANW18} likewise employs a single centralized critic, shared by all agents and conditioned on the joint action, again leaving no independent per-agent estimates whose divergence could signal miscoordination.
MADDPG~\cite{DBLP:conf/nips/LoweWTHAM17} is closer: each agent owns a critic $Q_i(s, a_1, \dots, a_N)$, but it is conditioned on the \emph{joint} continuous action, so the per-agent maximum $\max_{a_i} Q_i$ underlying the Q-value gap is not directly computable.
A DisQ variant for MADDPG-style architectures could instead compare the agents' critics evaluated at the same greedy joint action, $Q_i(s, \mu_1(o_1), \dots, \mu_N(o_N))$; developing and validating such a variant requires a different gap definition and calibration and is beyond the scope of this paper.

\paragraph{Access to Q-values.}
DisQ operates in a white-box (or at least gray-box) setting.
This is realistic when developers test their own systems before deployment, the setting we target: the Q-values are computed by the policies anyway, and exposing them requires no instrumentation beyond reading the network outputs.
DisQ is not applicable when third-party agents expose only their selected actions.
In such black-box settings, a possible direction is to train surrogate value functions from observed behavior, e.g., by distillation, and to apply DisQ to the surrogates. However, this would substantially increase the complexity and requirements for applying DisQ, making it an impractical alternative.

\paragraph{Initial-state Dependence.}
DisQ evaluates only initial states, so it cannot select for
miscoordination whose Q-value signal emerges later in an episode, or
for failures that are not accompanied by a Q-value gap at any step
(Section~\ref{sec:DisQ}).
RQ3.4 (Section~\ref{sec:midepisode}) probes this boundary empirically: it identifies an environment class in which DisQ yields no improvement and quantifies, per environment and training regime, how many failure episodes carry an initial-state gap, develop one only during execution, or never exhibit one.

\section{Experiments}\label{sec:experiments}
In this section, we evaluate DisQ in four environments and with different numbers of agents.
We structure the evaluation along four research questions:
\begin{itemize}
    \item[\textbf{RQ1}] \textbf{(Effectiveness)} To what extent does DisQ improve the failure-detection effectiveness of existing SBST methods?
    We compare random and genetic SBST with and without DisQ on the average episode reward of executed test cases and on the number, rate, and enrichment (odds ratio) of failure episodes per Definition~\ref{def:failure}.
    \item[\textbf{RQ2}] \textbf{(Gap--severity relation)} How does the magnitude of the Q-value gap relate to the severity of the exposed behavior?
    We sweep the selection quantile $p$ from 0 to 100 and measure the Spearman rank correlation between $p$ and the average episode reward of the selected test cases.
    \item[\textbf{RQ3}] \textbf{(Boundary conditions)} Under which variations of the CMARL setting does DisQ's advantage shrink or disappear?
    We vary the local-to-shared reward ratio (RQ3.1), the degree of initial-state cooperation required by the environment (RQ3.2), the training stage of the policies under test (RQ3.3), and the timing of Q-value divergence within episodes (RQ3.4), using the same reward and failure metrics as RQ1.
    \item[\textbf{RQ4}] \textbf{(Explainability)} What information does
    the Q-value gap provide for explaining a selected test case and its
    outcome? We inspect the per-agent Q-values of gap states selected
    during testing and relate the predicted returns to the observed
    episode rewards.
\end{itemize}
Before we answer each question individually, we introduce the used environments, trained CMARL policies, testing methods, and technical setup.

\subsection{Environments}
Multi-Agent Particle Environments (MPE) are a set of commonly used environments~\cite{DBLP:conf/issta/MaWWXWLX024} for CMARL benchmarking where particle agents can move, see each other, and interact with fixed landmarks~\cite{terry2021pettingzoo}.
\emph{Simple Spread} has $N$ agents, $N$ landmarks (default $N=3$).
Agents must learn to cover all the landmarks while avoiding collisions.
All agents are globally penalized (shared reward) based on how far the closest agent is to each landmark (the sum of the minimum distances).
Locally, the agents are penalized if they collide with other agents (-1 for each collision).
\emph{Simple Speaker Listener} (Speaker Listener for short) has one agent that is the ``speaker'' and can speak but cannot move, while the other agent is the listener (cannot speak but must navigate to the correct landmark). The agents receive fully shared rewards.
We refer to PettingZoo~\cite{terry2021pettingzoo} for a detailed description of the environments.

We further introduce two new environments:
In \emph{Coin Collector}, agents must cooperate to collect as many coins as possible in a grid world in a given amount of time.
When a coin is collected, the agents get no penalty; otherwise, they get a shared penalty.

Additionally, we introduce the \emph{Knight Wizard Zombies} environment, which is a variant of the Knights Archers Zombies environment from PettingZoo~\cite{terry2021pettingzoo}.
We design this variant specifically to support the analysis of different cooperation levels in Section~\ref{sec:limits}.
This environment is a grid-based cooperative game where two agents (a knight and a wizard) work together to fend off three zombies on a grid.
Each agent can move in four directions, with the knight able to attack zombies in the same position and the wizard capable of healing the knight at the cost of mana points.
Zombies move randomly toward the far side of the grid and can attack the knight or wizard if they occupy the same position, with a certain probability of success.
The game concludes when the knight or wizard runs out of health points or if any zombie reaches the other side of the grid.
Agents receive shared rewards for defeating zombies (which are then respawned) and for successful healing actions, promoting teamwork to survive the zombie onslaught.

\subsection{Trained CMARL Policies}
We trained centralized and decentralized CMARL agents using the deep Q-learning algorithm~\cite{mnih2013playing}.
The centralized CMARL policies were trained for 100 million steps, while the decentralized CMARL policies were trained for 100,000 steps. This difference arises due to the distinct technical setups of the two CMARL algorithms.
Each policy is represented by a neural network with three hidden layers, each containing 512 neurons.
Across 1,000 episodes, the trained centralized policies achieve an average return of approximately $-55$, while the decentralized policies achieve an average return of approximately $-32$.
As a robustness measure comparable across environments, we report each policy's failure rate under uniformly random initial states (Definition~\ref{def:failure} with the parametric threshold $\theta = \mu - 2\sigma$).
Over $25{,}000$ sampled episodes per environment, the policies fail in $0.0$--$4.7\%$ (centralized) and $0.0$--$5.6\%$ (decentralized) of episodes.
Coin Collector policies are the most robust (${\le}\,0.04\%$) and Speaker Listener policies the least.
The trained agents are thus strong enough that failures are rare events worth searching for.
For more details about the training process, we refer to the source code.

\subsection{Testing Methods}
We compare random SBST with and without DisQ and genetic SBST with and without DisQ.
The random SBST randomly samples test cases, while the genetic SBST optimizes the CMARL agent positions to minimize the collected reward.

\paragraph{Threshold Calibration}
Unless stated otherwise, DisQ-enhanced methods allocate the first $n = 25$ episodes of the test budget to calibrate $\lambda$ as the $p = 0.9$ quantile of the observed $D(s_0)$ values.
We chose $n = 25$ and $p = 0.9$ as a compromise between selectivity and throughput, and verified both empirically: since evaluating $D(s_0)$ only requires an environment reset and one forward pass per agent, we sampled $10^4$ initial states per trained policy as a reference distribution and bootstrapped the calibration $10^3$ times for every combination of $n \in \{25, 50, 100, 250\}$ and $p \in \{0.9, 0.95, 0.99\}$ (Table~\ref{tab:lambda_sensitivity}).
At $p = 0.9$, the threshold calibrated from $n = 25$ episodes deviates from the reference quantile by only $-3.9\% \pm 13.8\%$ on average, and increasing $n$ to 250 changes the mean calibrated threshold by less than 5\% for every policy; the achieved selection rate (the fraction of initial states that exceed $\hat{\lambda}$) stays close to the $10\%$ target.
Higher quantiles, in contrast, cannot be calibrated reliably within a small calibration budget: with $n = 25$ samples, at most one observation is expected above the 96th percentile, so at $p = 0.99$ the calibrated threshold is biased low by $15.7\%$ on average and the achieved selection rate varies between $0.4\%$ and $12.1\%$ (5th--95th percentile) instead of the targeted $1\%$.
The selection behavior then becomes uncontrolled, and the few episodes that are executed per test budget are too few to compare testing methods reliably.
RQ2 complements this analysis by varying $p$ over the full range and showing that higher quantiles indeed select test cases with lower average reward.

\begin{table}[t]
\centering
\caption{Sensitivity of the $\lambda$ calibration: bootstrapped relative error of $\hat{\lambda}$ (vs. the quantile of $10^4$ reference samples) and achieved selection rate ($\mathrm{P}(D(s_0) > \hat{\lambda})$, target $1-p$) with its 5th--95th percentile range, averaged over all eight trained policies (centralized and decentralized $\times$ four environments; $10^3$ bootstrap calibrations each).}
\label{tab:lambda_sensitivity}
\small
\begin{tabular}{rrrrl}
\toprule
$p$ & $n$ & Rel. err. $\hat{\lambda}$ & \multicolumn{2}{l}{Selection rate (target $1-p$)} \\
\midrule
0.90 & 25 & $-3.9\% \pm 13.8$ & 0.128 & [0.044, 0.246] \\
 & 50 & $-2.0\% \pm 10.2$ & 0.115 & [0.053, 0.194] \\
 & 100 & $-1.2\% \pm 7.2$ & 0.108 & [0.063, 0.161] \\
 & 250 & $-0.5\% \pm 4.7$ & 0.103 & [0.074, 0.137] \\
\midrule
0.95 & 25 & $-5.9\% \pm 14.8$ & 0.083 & [0.018, 0.182] \\
 & 50 & $-3.2\% \pm 11.0$ & 0.067 & [0.022, 0.128] \\
 & 100 & $-1.7\% \pm 8.3$ & 0.059 & [0.026, 0.101] \\
 & 250 & $-0.8\% \pm 5.3$ & 0.054 & [0.033, 0.079] \\
\midrule
0.99 & 25 & $-15.7\% \pm 15.0$ & 0.044 & [0.004, 0.121] \\
 & 50 & $-9.7\% \pm 12.5$ & 0.026 & [0.004, 0.068] \\
 & 100 & $-6.3\% \pm 11.2$ & 0.019 & [0.003, 0.046] \\
 & 250 & $-2.7\% \pm 7.7$ & 0.013 & [0.004, 0.026] \\
\bottomrule
\end{tabular}
\end{table}

\paragraph{Genetic Algorithm}
Genetic SBST with DisQ favors individuals (initial states) that yield low episode rewards.
We assign individuals without an initial-state Q-value gap the worst possible fitness, thereby removing them from the selection.
Each individual encodes the initial positions of the agents or entities in the environment.
An initial population is sampled uniformly at random from the valid state bounds.
After a population has been evaluated, the two best individuals are selected as parents.
A new population is generated by uniform per-gene crossover followed by bounded mutation.
See Table~\ref{tab:ga_hyperparameters} for the hyperparameter settings of the genetic algorithm.

\begin{table}[t]
\centering
\caption{Genetic algorithm hyperparameters.}
\label{tab:ga_hyperparameters}
\resizebox{\textwidth}{!}{%
\begin{tabular}{ll}
\toprule
\textbf{Hyperparameter} & \textbf{Value / Description} \\
\midrule
Population size & $200$ individuals \\
Initialization & Uniform random sampling within valid initial-state bounds \\
Fitness & Episode reward, minimized \\
Selection & Two best individuals in the current population \\
Crossover & Uniform per-gene crossover \\
Crossover rate & $0.5$ \\
Mutation & Independent bounded perturbation of each gene \\
Mutation rate & $0.1$ per gene \\
Replacement & Full generational replacement with offspring from selected parents \\
Invalid episode fitness & $\infty$ \\
\bottomrule
\end{tabular}}
\end{table}

\subsection{Technical Setup}
We ran all experiments on a system equipped with dual AMD Epyc 7763 processors, 2\,TB of RAM, NVIDIA A100 GPUs, and 4\,TB of NVMe storage.
We use stable-baselines3~\cite{stable-baselines3} for training the centralized CMARL policies, PyTorch~\cite{pytorch} for the decentralized CMARL policies, PettingZoo~\cite{terry2021pettingzoo} for the CMARL environments, and Gimitest~\cite{gimitest} to implement our testing methods.

\subsection{Evaluation Metrics and Statistical Analysis}\label{sec:metrics}
We assess the testing methods with two complementary metrics.
First, the \emph{average episode reward} over all executed test cases: a more effective testing method selects or generates episodes in which the agents perform worse, i.e., it achieves a lower average reward.
We report the average reward per method together with its relative change over the \emph{Random Testing} baseline.
Second, following the failure notion of Definition~\ref{def:failure}, we count \emph{discrete failures}.
Since the environments have widely different reward scales, we instantiate the threshold $\theta$ of Definition~\ref{def:failure} non-parametrically per policy under test: an executed episode counts as a failure if its reward is at or below the 5th percentile of the Random Testing reward distribution for the same policy.
By construction, Random Testing then has a failure rate of approximately $0.05$, and the other methods are compared by how strongly they enrich such failure episodes among the episodes they actually execute.
The rate is approximate rather than exact because return distributions are discrete in some environments: where many episodes share the return at the threshold, the whole tied block is labeled, which raises the baseline rate above $5\%$. 
This is most pronounced in Coin Collector, at $8.6\%$ centrally.
We report the absolute number of failures, the number of executed episodes, and the resulting failure rate.
Note that the DisQ-enhanced methods execute only the selected test cases and therefore run far fewer episodes than their vanilla counterparts within the same budget.

All statistical comparisons follow the guidelines of Arcuri and Briand~\cite{DBLP:journals/stvr/ArcuriB14} for assessing randomized algorithms.
The unit of analysis is one repetition: each testing experiment is repeated 25 times, and each repetition contributes its own average reward, since episode rewards within a repetition are not independent samples of the testing method's behavior.
For the reward metric, we first apply a Kruskal--Wallis omnibus test across all methods per policy under test, followed by post-hoc two-sided Mann--Whitney U tests of each method against the Random Testing baseline; a comparison is only called significant if the omnibus test rejects as well.
We report the Vargha--Delaney $\hat{A}_{12}$ effect size~\cite{vargha2000critique}, the probability that a repetition of the method yields a higher average reward than a repetition of the baseline: $\hat{A}_{12} < 0.5$ means the method uncovers lower rewards (i.e., is the more effective testing method), $\hat{A}_{12} = 0.5$ means no difference.
For the discrete failure counts, we compare each method against the baseline with Fisher's exact test on the $2 \times 2$ table of failure versus non-failure episodes and report the odds ratio (OR) as effect size, where $\mathrm{OR} > 1$ means the method's executed episodes are more likely to be failures.
Within each policy under test, the $p$-values of each family of comparisons are adjusted for multiple comparisons with the Holm--Bonferroni procedure.
We use a significance level of $\alpha = 0.05$ and mark adjusted $p$-values as $^{*}$ ($p < 0.05$), $^{**}$ ($p < 0.01$), $^{***}$ ($p < 0.001$), or n.s.\ (not significant).

\section{Results}\label{sec:results}

\subsection{RQ1: To what extent does DisQ improve the failure-detection effectiveness of existing SBST methods?}
\paragraph{Setup} In this experiment, we enhance both random and genetic SBST using the DisQ test case selection method.
These enhanced methods are then compared to their vanilla counterparts: random and genetic SBST.

\paragraph{Execution} We conducted experiments for each testing method using a test budget of 1,000 episodes.
To account for randomness, each testing experiment was repeated 25 times; the repetitions are the unit of the statistical analysis described in Section~\ref{sec:metrics}.

\begin{figure*}[t]
    \centering
    \includegraphics[width=\linewidth]{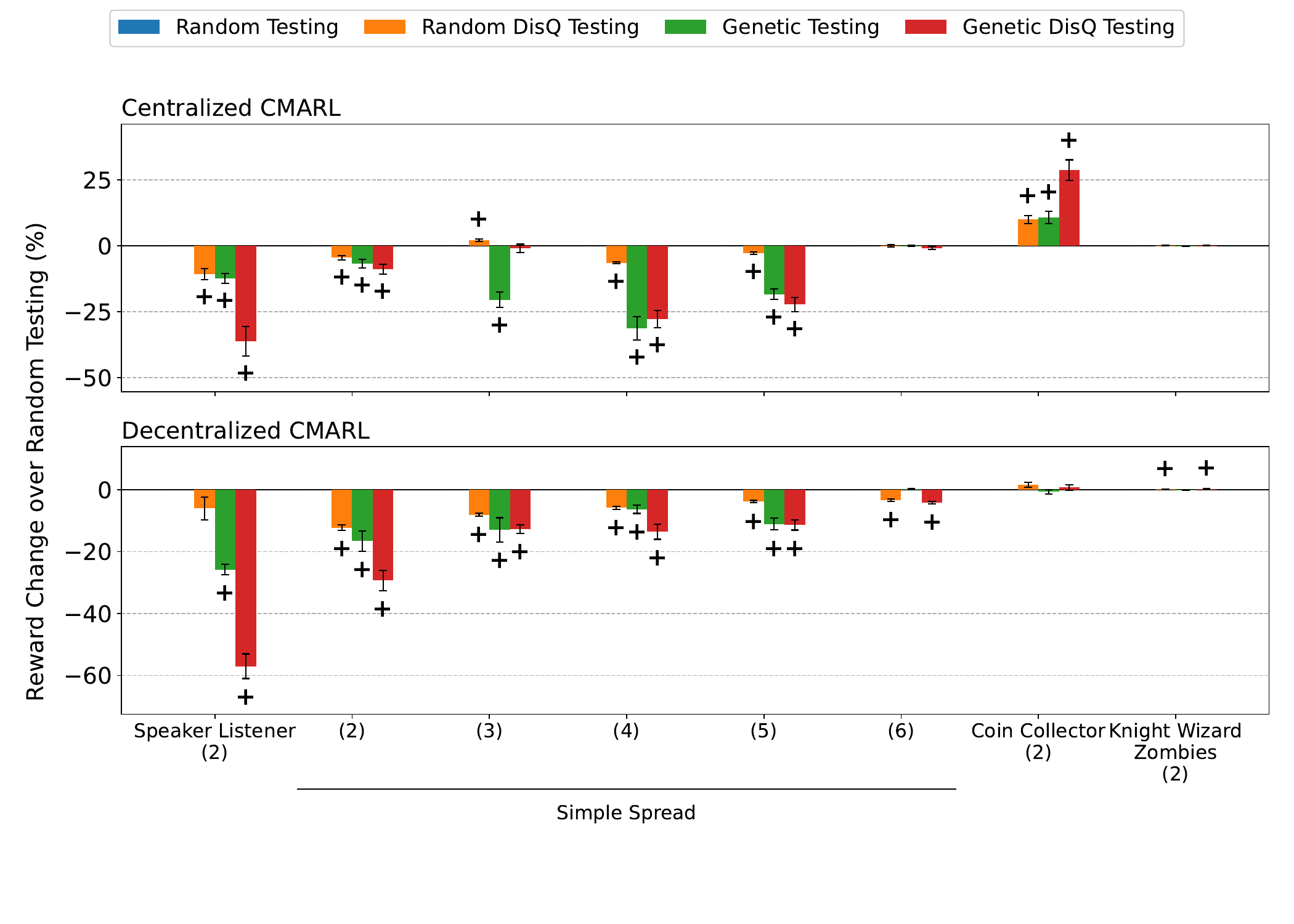}
    \caption{Relative change in average reward over Random Testing for the different testing methods, for centralized (top) and decentralized (bottom) CMARL, across all evaluated environments. The enhanced testing methods with DisQ outperform their vanilla alternatives in most cases. A `+' marks a statistically significant difference from Random Testing (same test and significance decision as Tables~\ref{tab:rq1_central} and~\ref{tab:rq1_decentral}); unmarked bars are not significant.}
    \label{fig:different_testing_methods}
\end{figure*}

\paragraph{Results}
Tables~\ref{tab:rq1_central} and~\ref{tab:rq1_decentral} report, for every environment, the average reward with its relative change over Random Testing, the $\hat{A}_{12}$ effect size and Holm-adjusted significance of the run-level Mann--Whitney comparison, and the discrete failure counts with their rates, odds ratios, and Fisher's exact significance.
Figure~\ref{fig:different_testing_methods} visualizes the relative reward changes.

\begin{table*}[tp]
\centering
\caption{RQ1: DisQ vs. vanilla SBST, centralized CMARL (baseline config).}
\label{tab:rq1_central}
\small
\setlength{\tabcolsep}{4.5pt}
\resizebox{\textwidth}{!}{\begin{tabular}{llrrrlrrrl}
\toprule
& & \multicolumn{4}{c}{Reward} & \multicolumn{4}{c}{Failures} \\
\cmidrule(lr){3-6}\cmidrule(lr){7-10}
Environment & Method & Avg. & $\Delta$\% & $\hat{A}_{12}$ & Sig. & Fail./Exec. & Rate & OR & Sig. \\
\midrule
Coin Collector (2) & Random Testing & 21.41 & +0.0 & -- & -- & 2,145/25,000 & 0.086 & -- & -- \\
 & Random DisQ Testing & 23.56 & +10.0 & 0.91 & *** & 470/3,570 & 0.132 & 1.62 & *** \\
 & Genetic Testing & 23.71 & +10.7 & 0.84 & *** & 3,826/25,000 & 0.153 & 1.93 & *** \\
 & Genetic DisQ Testing & 27.55 & +28.7 & 0.93 & *** & 1,983/7,941 & 0.250 & 3.55 & *** \\
\midrule
Knight Wizard Zombies (2) & Random Testing & -446.71 & +0.0 & -- & -- & 1,307/25,000 & 0.052 & -- & -- \\
 & Random DisQ Testing & -446.36 & +0.1 & 0.59 & n.s. & 167/3,658 & 0.046 & 0.87 & n.s. \\
 & Genetic Testing & -447.05 & -0.1 & 0.29 & n.s. & 1,422/25,000 & 0.057 & 1.09 & n.s. \\
 & Genetic DisQ Testing & -446.42 & +0.1 & 0.46 & n.s. & 513/9,789 & 0.052 & 1.00 & n.s. \\
\midrule
Speaker Listener (2) & Random Testing & -34.80 & +0.0 & -- & -- & 1,250/25,000 & 0.050 & -- & -- \\
 & Random DisQ Testing & -38.56 & -10.8 & 0.14 & *** & 385/3,535 & 0.109 & 2.32 & *** \\
 & Genetic Testing & -39.12 & -12.4 & 0.09 & *** & 2,012/25,000 & 0.080 & 1.66 & *** \\
 & Genetic DisQ Testing & -47.40 & -36.2 & 0.11 & *** & 619/3,395 & 0.182 & 4.24 & *** \\
\midrule
Simple Spread (2) & Random Testing & -56.58 & +0.0 & -- & -- & 1,250/25,000 & 0.050 & -- & -- \\
 & Random DisQ Testing & -59.11 & -4.5 & 0.09 & *** & 206/3,720 & 0.055 & 1.11 & n.s. \\
 & Genetic Testing & -60.42 & -6.8 & 0.16 & *** & 2,186/24,999 & 0.087 & 1.82 & *** \\
 & Genetic DisQ Testing & -61.55 & -8.8 & 0.10 & *** & 373/4,918 & 0.076 & 1.56 & *** \\
\midrule
Simple Spread (3) & Random Testing & -127.86 & +0.0 & -- & -- & 1,250/25,000 & 0.050 & -- & -- \\
 & Random DisQ Testing & -125.12 & +2.1 & 0.80 & *** & 138/4,026 & 0.034 & 0.67 & *** \\
 & Genetic Testing & -154.12 & -20.5 & 0.08 & *** & 5,166/24,987 & 0.207 & 4.95 & *** \\
 & Genetic DisQ Testing & -129.11 & -1.0 & 0.52 & n.s. & 215/3,648 & 0.059 & 1.19 & * \\
\midrule
Simple Spread (4) & Random Testing & -214.31 & +0.0 & -- & -- & 1,250/25,000 & 0.050 & -- & -- \\
 & Random DisQ Testing & -228.18 & -6.5 & 0.00 & *** & 241/3,465 & 0.070 & 1.42 & *** \\
 & Genetic Testing & -281.44 & -31.3 & 0.00 & *** & 6,795/24,995 & 0.272 & 7.09 & *** \\
 & Genetic DisQ Testing & -273.82 & -27.8 & 0.04 & *** & 1,318/4,998 & 0.264 & 6.80 & *** \\
\midrule
Simple Spread (5) & Random Testing & -313.84 & +0.0 & -- & -- & 1,250/25,000 & 0.050 & -- & -- \\
 & Random DisQ Testing & -322.44 & -2.7 & 0.10 & *** & 185/2,727 & 0.068 & 1.38 & *** \\
 & Genetic Testing & -371.66 & -18.4 & 0.00 & *** & 4,800/24,995 & 0.192 & 4.52 & *** \\
 & Genetic DisQ Testing & -383.80 & -22.3 & 0.07 & *** & 1,547/7,044 & 0.220 & 5.35 & *** \\
\midrule
Simple Spread (6) & Random Testing & -414.95 & +0.0 & -- & -- & 1,250/25,000 & 0.050 & -- & -- \\
 & Random DisQ Testing & -414.89 & +0.0 & 0.49 & n.s. & 231/4,192 & 0.055 & 1.11 & n.s. \\
 & Genetic Testing & -415.23 & -0.1 & 0.48 & n.s. & 1,297/25,000 & 0.052 & 1.04 & n.s. \\
 & Genetic DisQ Testing & -418.52 & -0.9 & 0.39 & n.s. & 184/3,353 & 0.055 & 1.10 & n.s. \\
\bottomrule
\end{tabular}}
\end{table*}

\begin{table*}[tp]
\centering
\caption{RQ1: DisQ vs. vanilla SBST, decentralized CMARL (baseline config).}
\label{tab:rq1_decentral}
\small
\setlength{\tabcolsep}{4.5pt}
\resizebox{\textwidth}{!}{\begin{tabular}{llrrrlrrrl}
\toprule
& & \multicolumn{4}{c}{Reward} & \multicolumn{4}{c}{Failures} \\
\cmidrule(lr){3-6}\cmidrule(lr){7-10}
Environment & Method & Avg. & $\Delta$\% & $\hat{A}_{12}$ & Sig. & Fail./Exec. & Rate & OR & Sig. \\
\midrule
Coin Collector (2) & Random Testing & 93.84 & +0.0 & -- & -- & 1,391/25,000 & 0.056 & -- & -- \\
 & Random DisQ Testing & 95.36 & +1.6 & 0.65 & n.s. & 203/3,920 & 0.052 & 0.93 & n.s. \\
 & Genetic Testing & 93.20 & -0.7 & 0.45 & n.s. & 1,677/25,000 & 0.067 & 1.22 & *** \\
 & Genetic DisQ Testing & 94.55 & +0.8 & 0.62 & n.s. & 188/3,813 & 0.049 & 0.88 & n.s. \\
\midrule
Knight Wizard Zombies (2) & Random Testing & -451.03 & +0.0 & -- & -- & 1,481/25,000 & 0.059 & -- & -- \\
 & Random DisQ Testing & -450.31 & +0.2 & 0.77 & ** & 216/3,673 & 0.059 & 0.99 & n.s. \\
 & Genetic Testing & -451.34 & -0.1 & 0.36 & n.s. & 1,589/25,000 & 0.064 & 1.08 & n.s. \\
 & Genetic DisQ Testing & -449.83 & +0.3 & 0.87 & *** & 433/7,875 & 0.055 & 0.92 & n.s. \\
\midrule
Speaker Listener (2) & Random Testing & -48.07 & +0.0 & -- & -- & 1,250/25,000 & 0.050 & -- & -- \\
 & Random DisQ Testing & -50.97 & -6.0 & 0.39 & n.s. & 491/4,310 & 0.114 & 2.44 & *** \\
 & Genetic Testing & -60.48 & -25.8 & 0.03 & *** & 2,691/25,000 & 0.108 & 2.29 & *** \\
 & Genetic DisQ Testing & -75.48 & -57.0 & 0.00 & *** & 1,008/4,444 & 0.227 & 5.57 & *** \\
\midrule
Simple Spread (2) & Random Testing & -34.51 & +0.0 & -- & -- & 1,250/25,000 & 0.050 & -- & -- \\
 & Random DisQ Testing & -38.74 & -12.3 & 0.00 & *** & 279/3,292 & 0.085 & 1.76 & *** \\
 & Genetic Testing & -40.23 & -16.6 & 0.21 & *** & 3,751/24,999 & 0.150 & 3.35 & *** \\
 & Genetic DisQ Testing & -44.63 & -29.3 & 0.00 & *** & 812/4,001 & 0.203 & 4.84 & *** \\
\midrule
Simple Spread (3) & Random Testing & -103.72 & +0.0 & -- & -- & 1,250/25,000 & 0.050 & -- & -- \\
 & Random DisQ Testing & -112.09 & -8.1 & 0.00 & *** & 309/3,663 & 0.084 & 1.75 & *** \\
 & Genetic Testing & -117.15 & -12.9 & 0.20 & *** & 3,238/25,000 & 0.130 & 2.83 & *** \\
 & Genetic DisQ Testing & -116.98 & -12.8 & 0.00 & *** & 507/4,348 & 0.117 & 2.51 & *** \\
\midrule
Simple Spread (4) & Random Testing & -184.93 & +0.0 & -- & -- & 1,250/25,000 & 0.050 & -- & -- \\
 & Random DisQ Testing & -195.74 & -5.8 & 0.01 & *** & 272/3,586 & 0.076 & 1.56 & *** \\
 & Genetic Testing & -196.52 & -6.3 & 0.11 & *** & 2,691/24,999 & 0.108 & 2.29 & *** \\
 & Genetic DisQ Testing & -210.11 & -13.6 & 0.00 & *** & 670/4,457 & 0.150 & 3.36 & *** \\
\midrule
Simple Spread (5) & Random Testing & -282.34 & +0.0 & -- & -- & 1,250/25,000 & 0.050 & -- & -- \\
 & Random DisQ Testing & -293.24 & -3.9 & 0.01 & *** & 261/3,942 & 0.066 & 1.35 & *** \\
 & Genetic Testing & -313.57 & -11.1 & 0.09 & *** & 4,074/25,000 & 0.163 & 3.70 & *** \\
 & Genetic DisQ Testing & -314.52 & -11.4 & 0.00 & *** & 610/4,255 & 0.143 & 3.18 & *** \\
\midrule
Simple Spread (6) & Random Testing & -396.37 & +0.0 & -- & -- & 1,250/25,000 & 0.050 & -- & -- \\
 & Random DisQ Testing & -409.82 & -3.4 & 0.02 & *** & 300/4,353 & 0.069 & 1.41 & *** \\
 & Genetic Testing & -395.75 & +0.2 & 0.59 & n.s. & 1,256/25,000 & 0.050 & 1.01 & n.s. \\
 & Genetic DisQ Testing & -413.07 & -4.2 & 0.03 & *** & 216/3,242 & 0.067 & 1.36 & *** \\
\bottomrule
\end{tabular}}
\end{table*}

Overall, the DisQ-enhanced methods detect worse behavior (lower rewards, more failures) than Random Testing in most environments, and in most cases at least match their vanilla counterparts.
In the decentralized setting, \emph{Genetic DisQ Testing} significantly reduces the average reward in six of the eight environment configurations (all but Coin Collector and Knight Wizard Zombies), with large effect sizes ($\hat{A}_{12} \le 0.04$) and failure odds ratios between $1.36$ and $5.57$; in five of these six environments it also reaches a lower average reward than vanilla \emph{Genetic Testing}.
In the centralized setting, the picture is similar for Speaker Listener and Simple Spread with 2, 4, and 5 agents (reward reductions of $8$--$36\%$, all $p < 0.001$; failure odds ratios up to $6.80$), with the strongest single result in Speaker Listener, where \emph{Genetic DisQ Testing} reduces the average reward by $36.2\%$ and more than triples the failure rate ($0.182$ vs.\ $0.050$, $\mathrm{OR} = 4.24$, $p < 0.001$).

The failure columns also expose an efficiency dimension that the average reward hides: because DisQ executes only the selected test cases, the DisQ methods run between $2{,}727$ and $9{,}789$ of the $25{,}000$ episodes (median $16\%$).
For example, in the decentralized Speaker Listener environment, \emph{Genetic DisQ Testing} finds $1{,}008$ failures in $4{,}444$ executed episodes, a similar absolute number to the $1{,}250$ failures Random Testing finds, but with less than a fifth of the execution effort.

Three deviations from this pattern are worth noting.
First, in the centralized Coin Collector environment, every non-baseline method \emph{increases} the average reward ($+10.0\%$ to $+28.7\%$, $\hat{A}_{12} \ge 0.84$) yet simultaneously and significantly \emph{increases} the failure rate ($0.132$ to $0.250$ vs.\ $0.086$, $\mathrm{OR}$ between $1.62$ and $3.55$, $p < 0.001$): the selected episodes are more extreme in both directions, and the discrete failure metric shows failure-revealing behavior that the average reward masks.
This affects vanilla \emph{Genetic Testing} as much as the DisQ variants, which points to the test generator rather than the selection method (Section~\ref{sec:threats}).
In the decentralized Coin Collector environment, only vanilla \emph{Genetic Testing} differs significantly from the baseline, and only on the failure metric ($0.067$ vs.\ $0.056$, $\mathrm{OR} = 1.22$); the initial state constrains the outcome only weakly in this environment, so a criterion evaluated at $s_0$ has little to act on.
Second, in centralized Simple Spread with 3 agents, \emph{Random DisQ Testing} selects episodes with significantly higher rewards and fewer failures ($\mathrm{OR} = 0.67$), and vanilla \emph{Genetic Testing} is the most effective method; DisQ is thus not uniformly beneficial.
Third, in Knight Wizard Zombies, no method achieves a practically relevant difference in either setting: all reward changes are at most $0.3\%$ in magnitude (even where statistically detectable), and no failure comparison is significant; we discuss this environment in detail in Section~\ref{sec:limits}.

\begin{figure*}[t]
    \centering
    \begin{subfigure}[b]{0.49\linewidth}
        \centering
        \includegraphics[width=\linewidth]{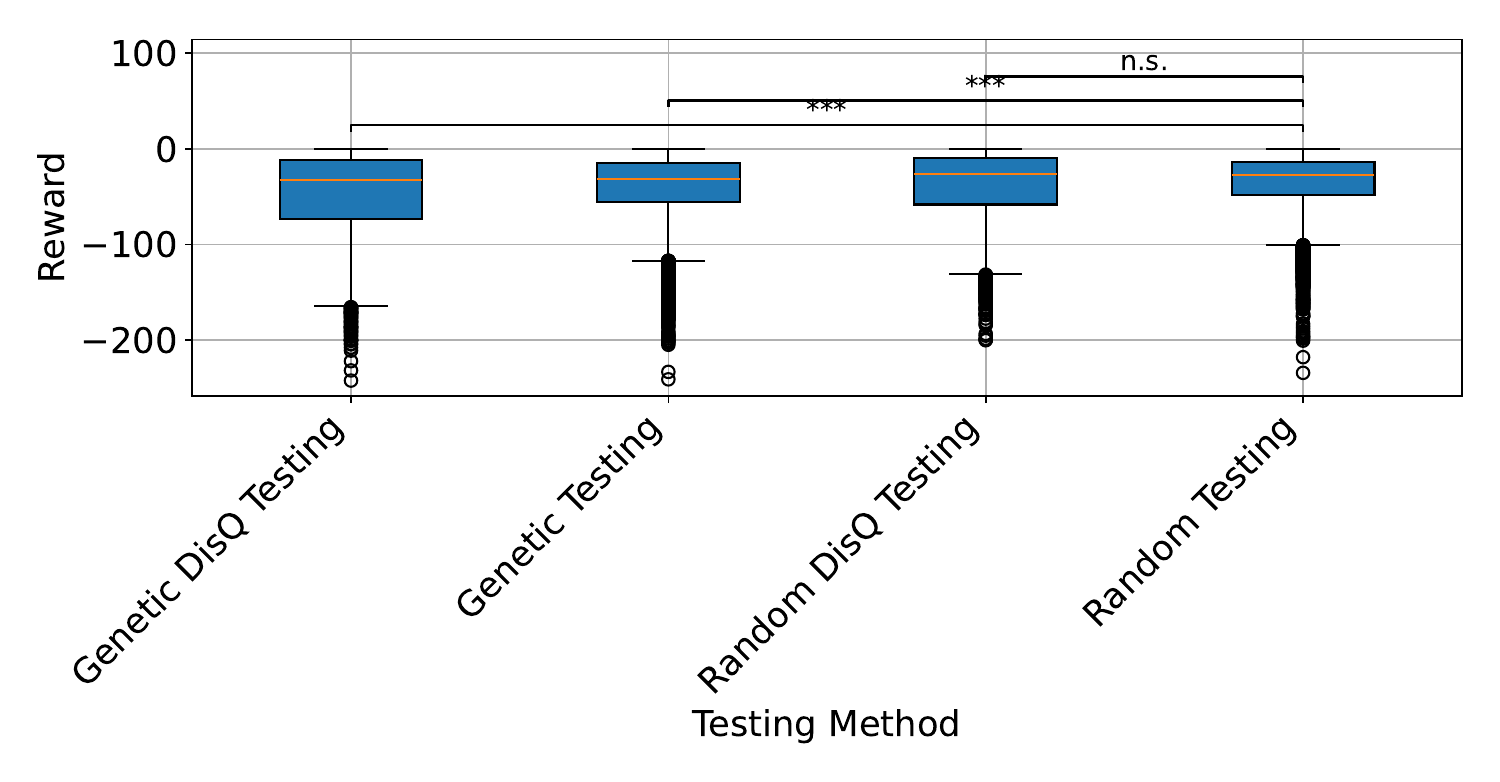}
        \caption{Centralized CMARL.}
    \end{subfigure}
    \hfill
    \begin{subfigure}[b]{0.49\linewidth}
        \centering
        \includegraphics[width=\linewidth]{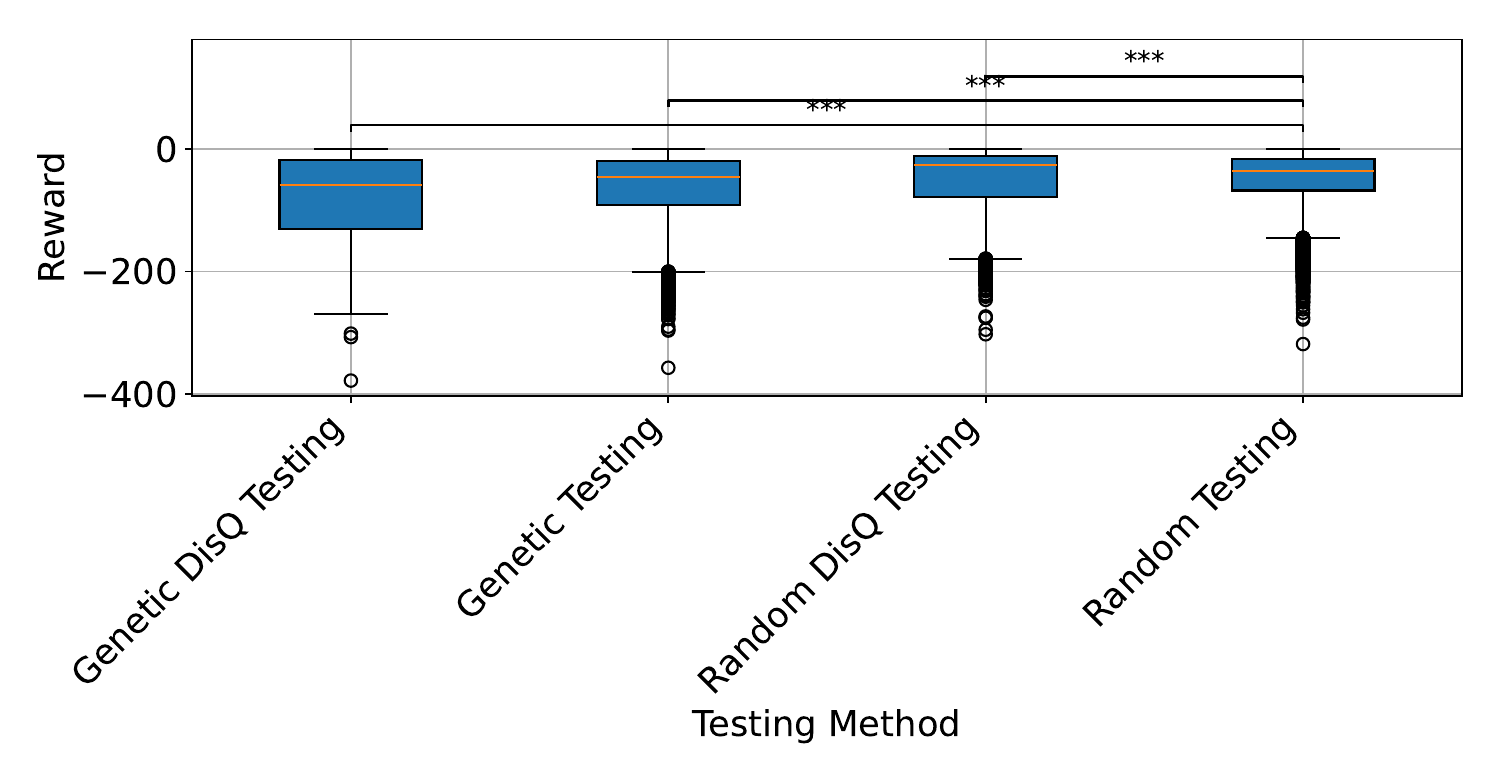}
        \caption{Decentralized CMARL.}
        \label{fig:dec_testing}
    \end{subfigure}
    \caption{Diagrams that visualize the rewards in the Simple Speaker Listener environment for the different testing methods in centralized and decentralized CMARL. We observe that the boxplots of the DisQ enhanced testing methods indicate that DisQ finds failure behavior more often.}
    \label{fig:different_testing_method_box}
\end{figure*}

In Figure~\ref{fig:different_testing_method_box}, we present the reward distributions for the Simple Speaker Listener environment across different testing methods. The box plots reveal how the episode rewards decline when comparing the enhanced testing methods to their vanilla~counterparts.

\paragraph{Answer}
DisQ improves both SBST methods on both metrics in the majority of environment--setting combinations.
The selected test cases reduce the average reward by up to $36.2\%$ over Random Testing ($\hat{A}_{12} \le 0.11$, Holm-adjusted $p < 0.001$) and enrich failure episodes by significant odds ratios between $1.19$ and $6.80$, while executing only $11$--$39\%$ of the episode budget.
The improvement is not universal: in centralized Simple Spread (3) Random DisQ selects fewer failures than the baseline ($\mathrm{OR} = 0.67$), and in Knight Wizard Zombies no method changes the reward by more than $0.3\%$.
Section~\ref{sec:limits} examines the conditions under which the advantage disappears.

\subsection{RQ2: How does the magnitude of the Q-value gap relate to the severity of the exposed behavior?}\label{sec:worse}
\paragraph{Setup}
In this experiment, we quantify the relation between the Q-value gap magnitude and the severity of the exposed behavior, measured as the average accumulated episode reward of the selected test cases.
We conduct this sweep on Simple Spread (2), where DisQ is effective under both training regimes (RQ1), so that the effect of the quantile is not confounded by an environment in which the gap signal itself is weak.

\paragraph{Execution}
We conducted experiments for each DisQ-enhanced testing method for centralized and decentralized policies, using a test budget of 1,000 episodes across $\lambda$-values, determined by quantile percentages ranging from 0 to 100 (25 repetitions, as in RQ1).

\begin{figure*}[t]
    \centering
    \begin{subfigure}[b]{0.49\linewidth}
        \centering
        \includegraphics[width=\linewidth]{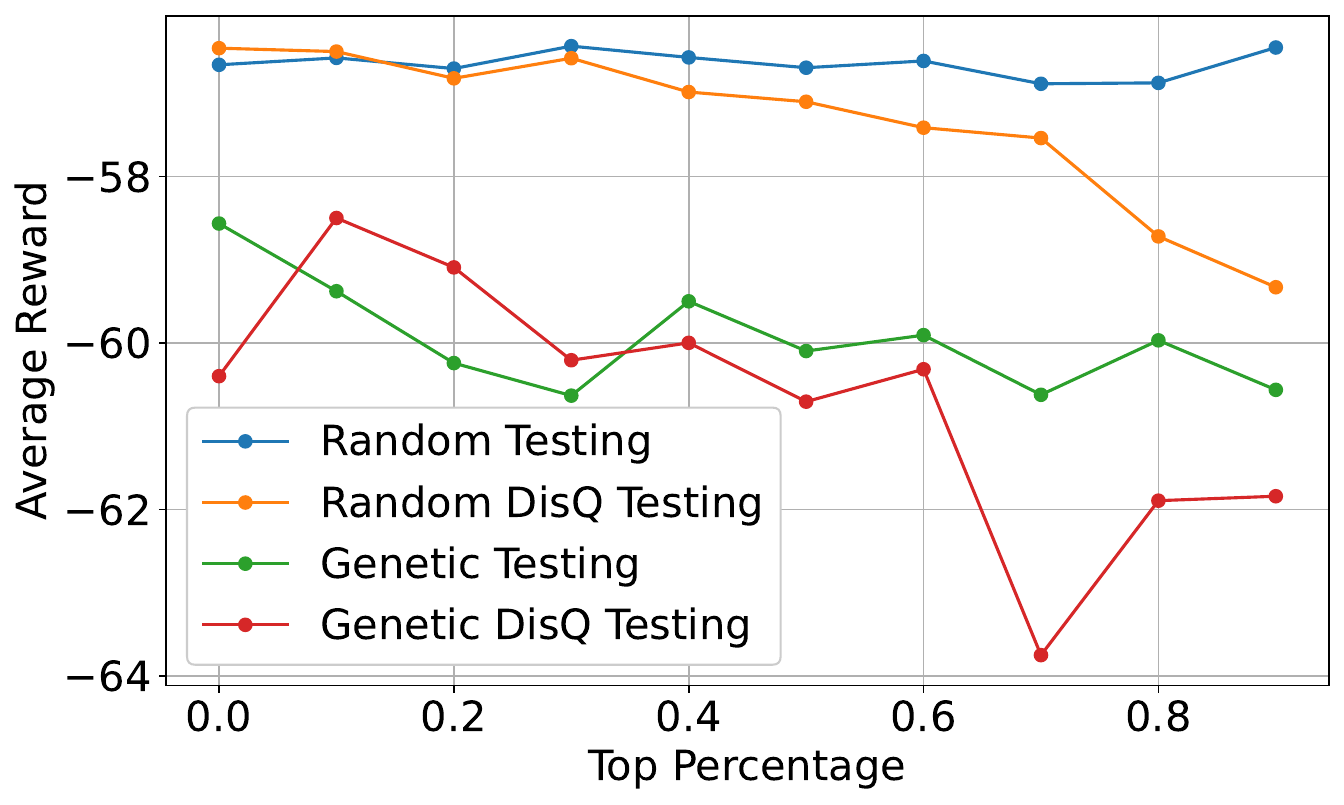}
        \caption{Centralized CMARL.}
    \end{subfigure}
    \hfill
    \begin{subfigure}[b]{0.49\linewidth}
        \centering
        \includegraphics[width=\linewidth]{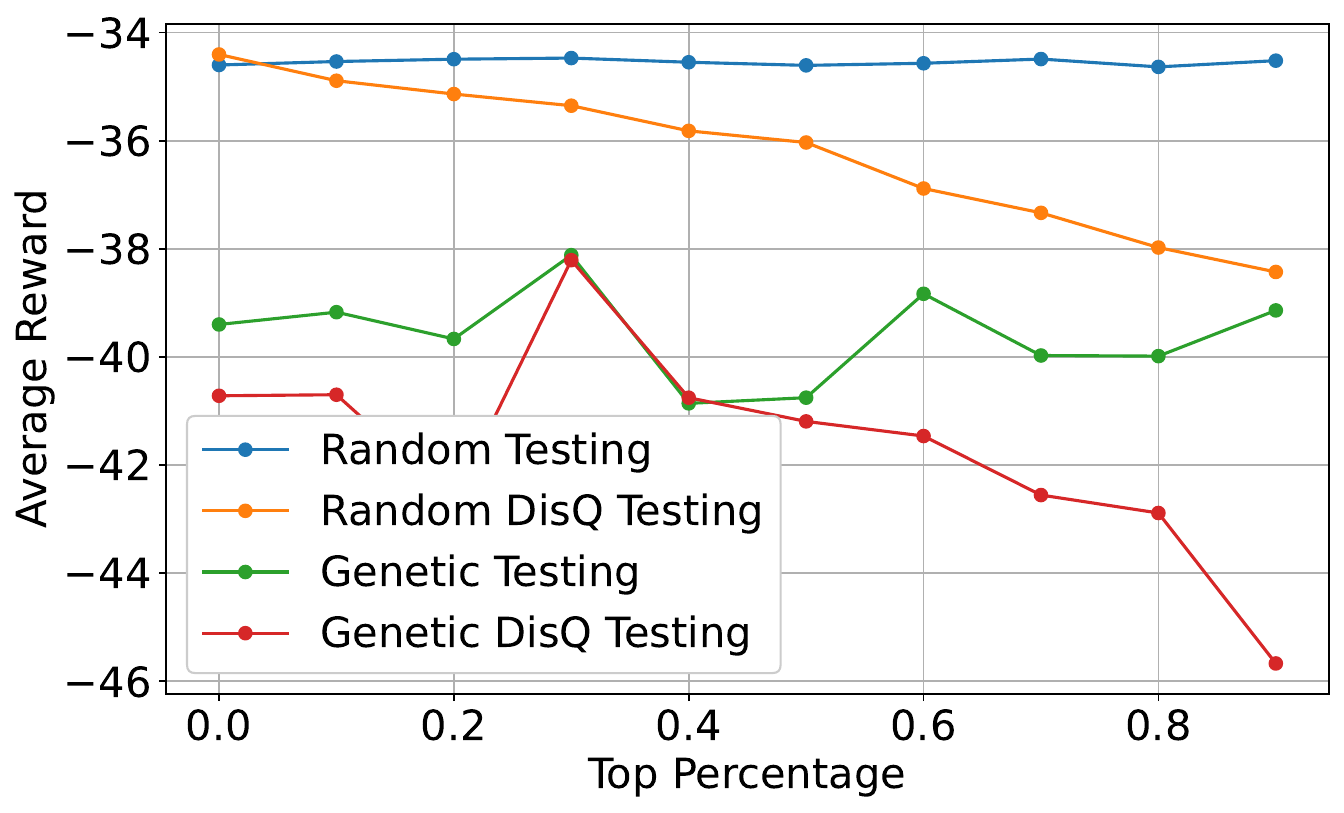}
        \caption{Decentralized CMARL.}
    \end{subfigure}
    \caption{Diagrams that visualize the average reward in the Simple Spread (2) environment over increasing q\%-quantiles (Top Percentage) for centralized and decentralized CMARL. We observed in both cases that the larger the Q-value gap, the lower the average reward, showing the effect of DisQ.}
    \label{fig:increased_threshold}
\end{figure*}

\paragraph{Results}
The results of the sweep, conducted on the Simple Spread (2) environment, are shown in Figure~\ref{fig:increased_threshold}; Table~\ref{tab:rq2_correlation} reports the Spearman rank correlation between the quantile percentage and the average reward for every method.
We observe that higher quantile percentages, i.e., higher Q-value gaps, are associated with lower average accumulated episode reward.
The correlation is strong and significant for the DisQ-guided methods: $\rho = -0.95$ (Random DisQ) and $\rho = -0.65$ (Genetic DisQ) for centralized policy testing, and $\rho = -1.00$ (Random DisQ) and $\rho = -0.82$ (Genetic DisQ) for decentralized policy testing (all $p < 0.05$).
For the vanilla testing methods, which do not use the selection threshold, no significant correlation exists ($|\rho| \le 0.60$, all $p > 0.06$): their variation arises solely from the randomness of the sampling process.

\begin{table}[t]
\centering
\caption{Spearman correlation between top-percentage (Q-distance quantile $\lambda$) and average reward, Simple Spread (2).}
\label{tab:rq2_correlation}
\begin{tabular}{llrr}
\toprule
Setting & Method & $\rho$ & $p$ \\
\midrule
Centralized & Genetic DisQ Testing & -0.65 & 4.25e-02 \\
 & Genetic Testing & 0.60 & 6.67e-02 \\
 & Random DisQ Testing & -0.95 & 2.28e-05 \\
 & Random Testing & -0.04 & 9.07e-01 \\
\midrule
Decentralized & Genetic DisQ Testing & -0.82 & 3.81e-03 \\
 & Genetic Testing & -0.16 & 6.51e-01 \\
 & Random DisQ Testing & -1.00 & 6.65e-64 \\
 & Random Testing & 0.39 & 2.60e-01 \\
\bottomrule
\end{tabular}
\end{table}

\paragraph{Answer}
Our results show that higher Q-value gaps correlate with worse behavior, i.e., larger reward reduction.
The selection quantile, and hence the Q-value gap magnitude, is strongly negatively rank-correlated with the average reward of the selected test cases: $\rho \in [-1.00, -0.65]$ for the DisQ-guided methods across both training regimes (all $p < 0.05$), while the vanilla methods show no significant correlation ($|\rho| \le 0.60$).
This means that DisQ is not only a go/no-go selection heuristic, but can also rank and prioritize test cases by failure severity.

\subsection{RQ3: Under which variations of the CMARL setting does DisQ's advantage shrink or disappear?}\label{sec:limits}
In this experiment, we examine the limitations of the DisQ test case selection by investigating local vs. shared rewards (RQ3.1), different cooperation levels in the CMARL environment (RQ3.2), different training stages of the CMARL policies under test (RQ3.3), and the divergence of Q-values mid-episode (RQ3.4).
RQ3.1 and RQ3.3 vary one factor at a time on Simple Spread (2), the reference environment in which DisQ is effective under both regimes (RQ1).
RQ3.2 and RQ3.4 contrast it with Knight Wizard Zombies, where it is not.

\subsubsection{RQ3.1: Ratio of Local to Shared Reward}

\paragraph{Setup}
In this experiment, we investigate if different percentages of local reward to shared reward influence the DisQ test case selection method.

\paragraph{Execution}
We conducted experiments for each DisQ-enhanced testing method for centralized and decentralized policies.
We used a test budget of 1,000 episodes across local ratios from all shared rewards (0) up to all local reward (1) (25 repetitions, as in RQ1).

\begin{figure*}[t]
    \centering
    \begin{subfigure}[b]{0.49\linewidth}
        \centering
        \includegraphics[width=\linewidth]{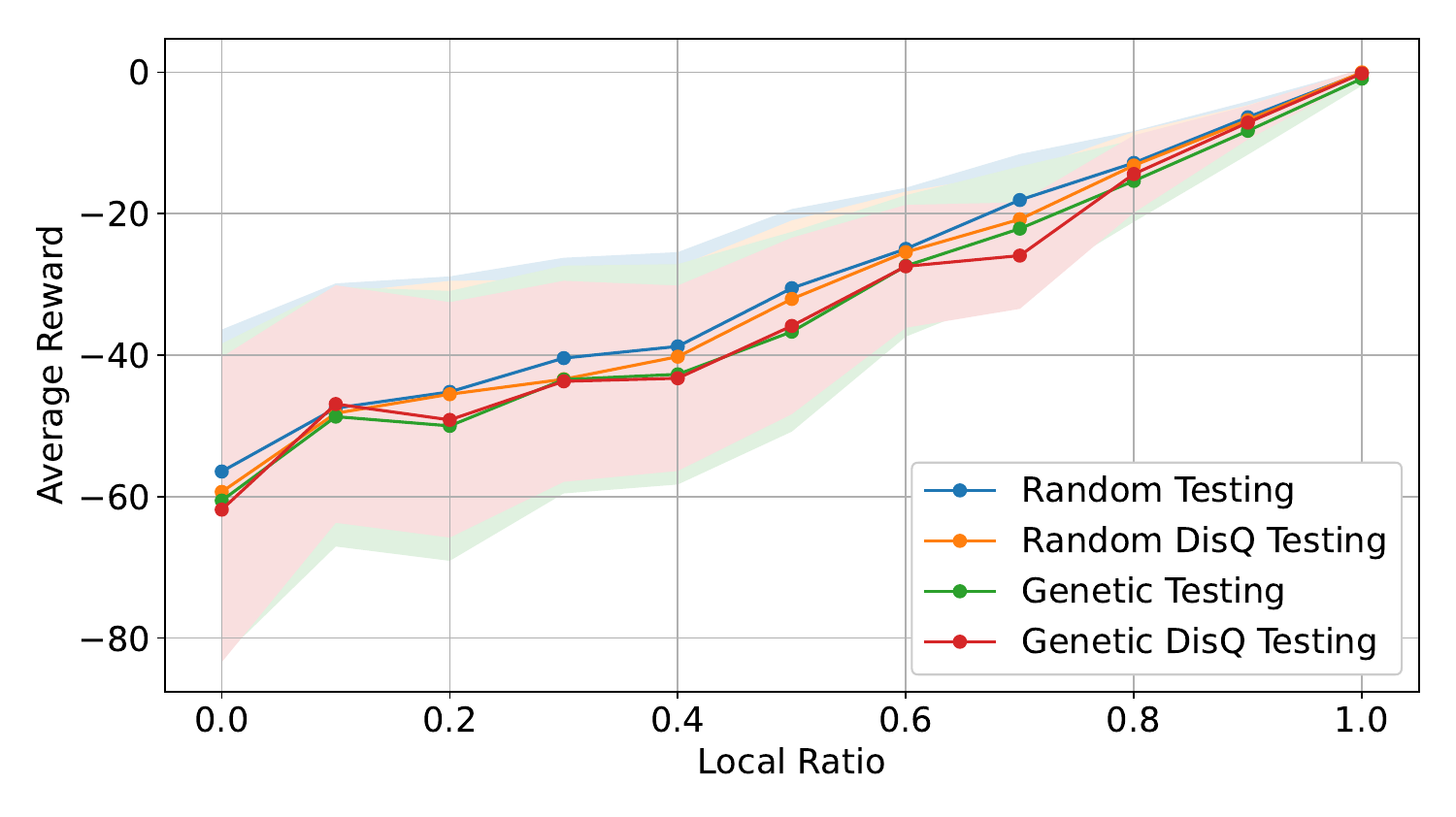}
        \caption{Centralized CMARL.}
    \end{subfigure}
    \hfill
    \begin{subfigure}[b]{0.49\linewidth}
        \centering
        \includegraphics[width=\linewidth]{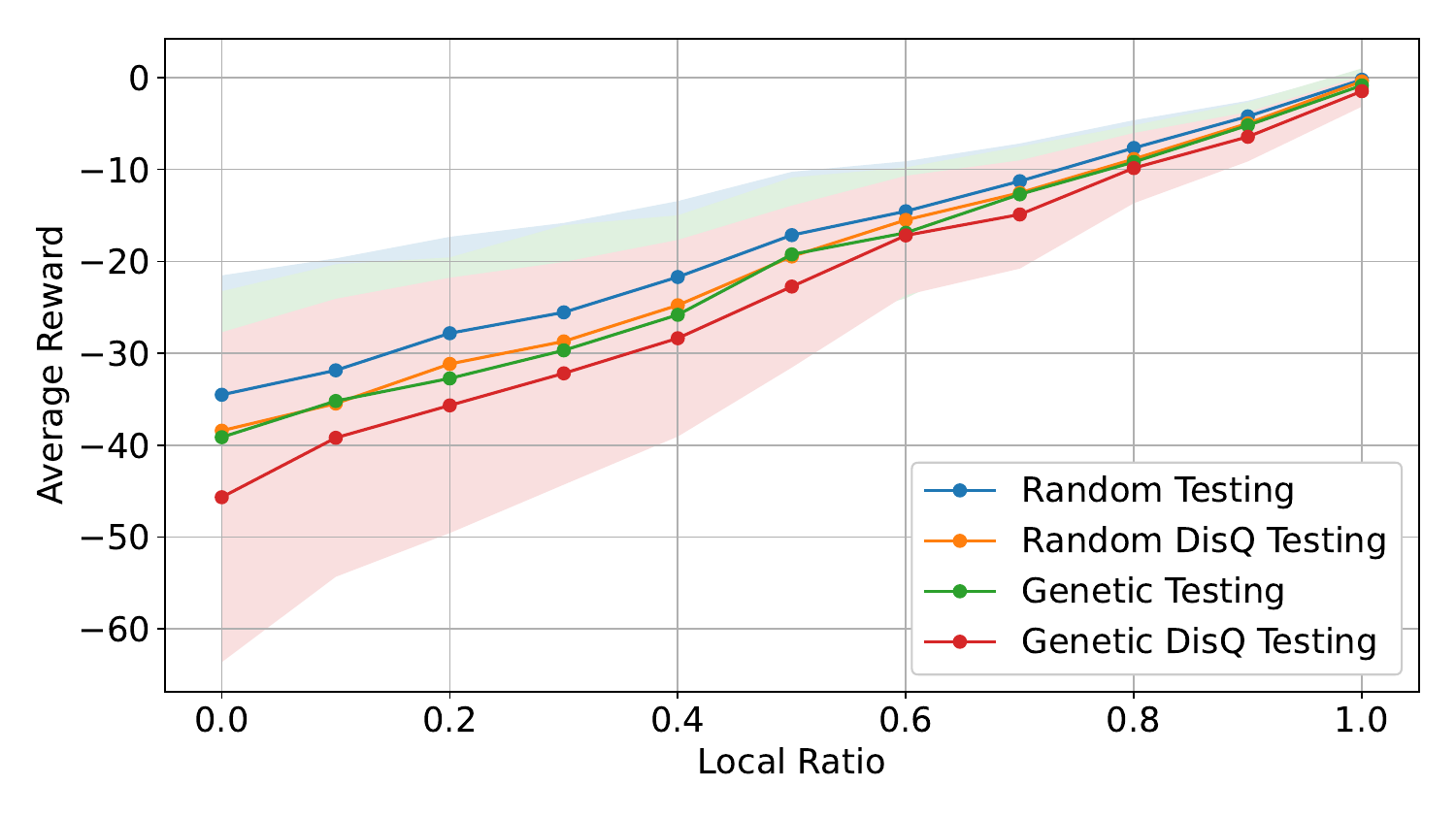}
        \caption{Decentralized CMARL.}
    \end{subfigure}
    \caption{Diagrams illustrating the effect of DisQ on the average reward in the Simple Spread (2) environment show a diminishing impact of DisQ as the local reward ratio increases. It also shows that DisQ outperforms the vanilla testing methods up to a local ratio of 0.9.}
    \label{fig:local_vs_global}
\end{figure*}

\paragraph{Result}
Figure~\ref{fig:local_vs_global} presents the testing results of the Simple Spread (2) environment, evaluated across varying percentages of local reward relative to shared reward; Table~\ref{tab:rq3_1_local_ratio} lists the underlying average rewards.
In the decentralized setting, \emph{Genetic DisQ Testing} achieves the lowest average reward at every ratio; in the centralized setting, the DisQ-enhanced variants stay below Random Testing up to a local ratio of $0.9$, but no single method dominates uniformly.
As the local ratio grows, the rewards of all methods converge towards zero and the absolute differences shrink.
The ordering is thus preserved across ratios, even though the general testing effectiveness decreases.

\begin{table*}[tp]
\centering
\caption{RQ3.1: Average reward vs. local reward ratio, Simple Spread (2).}
\label{tab:rq3_1_local_ratio}
\small
\resizebox{\textwidth}{!}{\begin{tabular}{llrrrrrrrrrrr}
\toprule
& Method & 0.0 & 0.1 & 0.2 & 0.3 & 0.4 & 0.5 & 0.6 & 0.7 & 0.8 & 0.9 & 1.0 \\
\midrule
Centralized & Random Testing & -56.5 & -47.5 & -45.2 & -40.4 & -38.8 & -30.5 & -25.0 & -18.1 & -12.8 & -6.4 & -0.1 \\
 & Random DisQ Testing & -59.3 & -48.2 & -45.5 & -43.4 & -40.2 & -32.1 & -25.4 & -20.8 & -13.2 & -6.8 & -0.0 \\
 & Genetic Testing & -60.6 & -48.7 & -50.0 & -43.4 & -42.7 & -36.7 & -27.4 & -22.1 & -15.4 & -8.3 & -0.9 \\
 & Genetic DisQ Testing & -61.8 & -46.9 & -49.2 & -43.7 & -43.3 & -35.9 & -27.4 & -25.9 & -14.4 & -7.1 & -0.2 \\
\midrule
Decentralized & Random Testing & -34.5 & -31.9 & -27.8 & -25.6 & -21.7 & -17.1 & -14.6 & -11.3 & -7.7 & -4.2 & -0.2 \\
 & Random DisQ Testing & -38.4 & -35.5 & -31.2 & -28.7 & -24.8 & -19.5 & -15.5 & -12.5 & -8.9 & -5.0 & -0.4 \\
 & Genetic Testing & -39.1 & -35.2 & -32.7 & -29.7 & -25.8 & -19.2 & -16.9 & -12.7 & -9.2 & -5.2 & -0.9 \\
 & Genetic DisQ Testing & -45.7 & -39.2 & -35.7 & -32.2 & -28.4 & -22.7 & -17.2 & -14.9 & -9.8 & -6.4 & -1.5 \\
\bottomrule
\end{tabular}}
\end{table*}

\subsubsection{RQ3.2: Different Cooperation Levels}\label{sec:cooplevels}
We hypothesize that DisQ is effective in environments where the initial agent collaboration directly impacts the overall episode performance.
Simple Spread and Simple Speaker Listener are of this kind.

\paragraph{Setup}
To further assess the limitations of DisQ, we evaluate it in a different context: the Knight Wizard Zombies environment.
In this environment, initial agent collaboration has a minimal impact on the accumulated reward.
Specifically, the knight is the primary reward collector, as it must approach the zombies.
The wizard's role is to maintain distance from the zombies and apply healing at the appropriate moments. At the start of the episode, the wizard's actions do not meaningfully influence the environment, reducing the need for early collaboration.

\paragraph{Execution}
We conducted experiments for each DisQ-enhanced testing method for centralized and decentralized policies, using a test budget of 1,000 episodes in the Knight Wizard Zombies environment (25 repetitions, as in RQ1).

\paragraph{Result}
We observe that DisQ does not improve testing effectiveness where the initial cooperation has little impact on the episode outcome. As the rightmost bars in Figure~\ref{fig:different_testing_methods} show, applying DisQ yields no improvement, but also no degradation.

This is consistent with the premise of the experiment: the initial agent positions account for $0.4\%$ of the episode-reward variance in this environment, against $7.0\%$ in Simple Spread (2) (Section~\ref{sec:threats}), leaving little for \emph{any} initial-state-based selection to exploit.

\subsubsection{RQ3.3: Different Training Stages}

\paragraph{Setup}
We investigate how different trained CMARL policies (the same architecture and hyperparameters trained for different numbers of steps) influence the performance of the DisQ test case selection method.

\paragraph{Execution}
We conducted experiments for each DisQ-enhanced testing method for centralized and decentralized policies.
We used a test budget of 1,000 episodes across differently trained CMARL policies with training steps up to 100 million (25 repetitions, as in RQ1).

\paragraph{Result}
Tables~\ref{tab:rq3_3_central} and~\ref{tab:rq3_3_decentral} report the results per training stage for the Simple Spread (2) scenario, with the same reward and failure statistics as in RQ1.
We observe that the DisQ test case selection is less effective for less-trained policies than for better-trained ones.
In the decentralized setting, there is a clear crossover: for policies trained for up to $4{,}000$ steps, vanilla \emph{Genetic Testing} detects lower rewards than \emph{Genetic DisQ Testing} (e.g., $-30.3\%$ vs.\ $-14.4\%$ at $4{,}000$ steps), whereas from $8{,}000$ training steps on, the DisQ-enhanced variant reaches lower average rewards and, in most stages, higher failure odds (e.g., $\mathrm{OR} = 5.28$ vs.\ $1.99$ at $32{,}000$ steps).
In the centralized setting, the dependence on training quality is even more pronounced: for intermediately trained policies, \emph{Random DisQ Testing} is ineffective or even counterproductive, selecting episodes with significantly \emph{higher} rewards and fewer failures than random sampling ($\mathrm{OR}$ down to $0.53$), and only for the fully trained policy does it significantly enrich failures again ($\mathrm{OR} = 1.58$, $p < 0.001$).
This is consistent with the mechanism behind DisQ: earlier in training the Q-values are still poor approximations of the expected returns, so the Q-value gap only becomes a reliable indicator of failure behavior once the value estimates themselves are trustworthy.

\begin{table*}[tp]
\centering
\caption{RQ3.3: Effect of training duration, Simple Spread (2), centralized CMARL.}
\label{tab:rq3_3_central}
\small
\setlength{\tabcolsep}{4.5pt}
\resizebox{\textwidth}{!}{\begin{tabular}{llrrrlrrrl}
\toprule
& & \multicolumn{4}{c}{Reward} & \multicolumn{4}{c}{Failures} \\
\cmidrule(lr){3-6}\cmidrule(lr){7-10}
Train episodes & Method & Avg. & $\Delta$\% & $\hat{A}_{12}$ & Sig. & Fail./Exec. & Rate & OR & Sig. \\
\midrule
100 & Random Testing & -166.67 & +0.0 & -- & -- & 1,250/25,000 & 0.050 & -- & -- \\
 & Random DisQ Testing & -174.58 & -4.8 & 0.06 & *** & 306/3,855 & 0.079 & 1.64 & *** \\
 & Genetic Testing & -237.32 & -42.4 & 0.00 & *** & 7,698/24,983 & 0.308 & 8.46 & *** \\
 & Genetic DisQ Testing & -248.32 & -49.0 & 0.00 & *** & 3,135/7,589 & 0.413 & 13.37 & *** \\
\midrule
1,000 & Random Testing & -190.84 & +0.0 & -- & -- & 1,250/25,000 & 0.050 & -- & -- \\
 & Random DisQ Testing & -179.67 & +5.9 & 1.00 & *** & 104/3,497 & 0.030 & 0.58 & *** \\
 & Genetic Testing & -245.09 & -28.4 & 0.00 & *** & 7,284/24,999 & 0.291 & 7.81 & *** \\
 & Genetic DisQ Testing & -214.69 & -12.5 & 0.31 & * & 878/5,477 & 0.160 & 3.63 & *** \\
\midrule
10,000 & Random Testing & -75.87 & +0.0 & -- & -- & 1,250/25,000 & 0.050 & -- & -- \\
 & Random DisQ Testing & -76.48 & -0.8 & 0.45 & n.s. & 171/3,922 & 0.044 & 0.87 & n.s. \\
 & Genetic Testing & -102.91 & -35.6 & 0.00 & *** & 6,784/25,000 & 0.271 & 7.08 & *** \\
 & Genetic DisQ Testing & -85.75 & -13.0 & 0.16 & *** & 679/5,989 & 0.113 & 2.43 & *** \\
\midrule
100,000 & Random Testing & -189.00 & +0.0 & -- & -- & 1,250/25,000 & 0.050 & -- & -- \\
 & Random DisQ Testing & -189.80 & -0.4 & 0.42 & n.s. & 150/3,913 & 0.038 & 0.76 & ** \\
 & Genetic Testing & -241.62 & -27.8 & 0.00 & *** & 7,885/24,997 & 0.315 & 8.75 & *** \\
 & Genetic DisQ Testing & -228.18 & -20.7 & 0.00 & *** & 698/3,857 & 0.181 & 4.20 & *** \\
\midrule
1,000,000 & Random Testing & -192.25 & +0.0 & -- & -- & 1,250/25,000 & 0.050 & -- & -- \\
 & Random DisQ Testing & -184.17 & +4.2 & 0.91 & *** & 133/3,466 & 0.038 & 0.76 & ** \\
 & Genetic Testing & -248.63 & -29.3 & 0.00 & *** & 7,889/24,999 & 0.316 & 8.76 & *** \\
 & Genetic DisQ Testing & -234.39 & -21.9 & 0.12 & *** & 1,206/5,556 & 0.217 & 5.27 & *** \\
\midrule
10,000,000 & Random Testing & -192.69 & +0.0 & -- & -- & 1,250/25,000 & 0.050 & -- & -- \\
 & Random DisQ Testing & -176.06 & +8.6 & 1.00 & *** & 104/3,809 & 0.027 & 0.53 & *** \\
 & Genetic Testing & -248.68 & -29.1 & 0.00 & *** & 7,555/24,998 & 0.302 & 8.23 & *** \\
 & Genetic DisQ Testing & -229.20 & -18.9 & 0.12 & *** & 1,202/6,034 & 0.199 & 4.73 & *** \\
\midrule
100,000,000 & Random Testing & -57.00 & +0.0 & -- & -- & 1,250/25,000 & 0.050 & -- & -- \\
 & Random DisQ Testing & -60.98 & -7.0 & 0.00 & *** & 293/3,818 & 0.077 & 1.58 & *** \\
 & Genetic Testing & -63.00 & -10.5 & 0.24 & ** & 2,864/25,000 & 0.115 & 2.46 & *** \\
 & Genetic DisQ Testing & -66.48 & -16.6 & 0.04 & *** & 467/3,818 & 0.122 & 2.65 & *** \\
\bottomrule
\end{tabular}}
\end{table*}

\begin{table*}[tp]
\centering
\caption{RQ3.3: Effect of training duration, Simple Spread (2), decentralized CMARL.}
\label{tab:rq3_3_decentral}
\small
\setlength{\tabcolsep}{4.5pt}
\resizebox{\textwidth}{!}{\begin{tabular}{llrrrlrrrl}
\toprule
& & \multicolumn{4}{c}{Reward} & \multicolumn{4}{c}{Failures} \\
\cmidrule(lr){3-6}\cmidrule(lr){7-10}
Train episodes & Method & Avg. & $\Delta$\% & $\hat{A}_{12}$ & Sig. & Fail./Exec. & Rate & OR & Sig. \\
\midrule
2,000 & Random Testing & -99.78 & +0.0 & -- & -- & 1,250/25,000 & 0.050 & -- & -- \\
 & Random DisQ Testing & -105.73 & -6.0 & 0.05 & *** & 257/3,615 & 0.071 & 1.45 & *** \\
 & Genetic Testing & -126.48 & -26.8 & 0.00 & *** & 5,098/25,000 & 0.204 & 4.87 & *** \\
 & Genetic DisQ Testing & -124.40 & -24.7 & 0.00 & *** & 1,086/6,977 & 0.156 & 3.50 & *** \\
\midrule
4,000 & Random Testing & -70.76 & +0.0 & -- & -- & 1,250/25,000 & 0.050 & -- & -- \\
 & Random DisQ Testing & -73.20 & -3.5 & 0.05 & *** & 200/3,827 & 0.052 & 1.05 & n.s. \\
 & Genetic Testing & -92.21 & -30.3 & 0.00 & *** & 6,412/25,000 & 0.256 & 6.55 & *** \\
 & Genetic DisQ Testing & -80.97 & -14.4 & 0.07 & *** & 593/4,638 & 0.128 & 2.79 & *** \\
\midrule
8,000 & Random Testing & -51.62 & +0.0 & -- & -- & 1,250/25,000 & 0.050 & -- & -- \\
 & Random DisQ Testing & -55.48 & -7.5 & 0.01 & *** & 275/3,840 & 0.072 & 1.47 & *** \\
 & Genetic Testing & -57.15 & -10.7 & 0.12 & *** & 2,703/25,000 & 0.108 & 2.30 & *** \\
 & Genetic DisQ Testing & -59.31 & -14.9 & 0.04 & *** & 528/4,464 & 0.118 & 2.55 & *** \\
\midrule
16,000 & Random Testing & -41.24 & +0.0 & -- & -- & 1,250/25,000 & 0.050 & -- & -- \\
 & Random DisQ Testing & -45.15 & -9.5 & 0.00 & *** & 274/3,944 & 0.069 & 1.42 & *** \\
 & Genetic Testing & -42.54 & -3.1 & 0.36 & n.s. & 1,702/25,000 & 0.068 & 1.39 & *** \\
 & Genetic DisQ Testing & -50.61 & -22.7 & 0.00 & *** & 776/5,363 & 0.145 & 3.21 & *** \\
\midrule
32,000 & Random Testing & -36.89 & +0.0 & -- & -- & 1,250/25,000 & 0.050 & -- & -- \\
 & Random DisQ Testing & -40.00 & -8.4 & 0.01 & *** & 261/3,786 & 0.069 & 1.41 & *** \\
 & Genetic Testing & -40.01 & -8.5 & 0.30 & * & 2,373/24,987 & 0.095 & 1.99 & *** \\
 & Genetic DisQ Testing & -49.15 & -33.2 & 0.00 & *** & 897/4,125 & 0.217 & 5.28 & *** \\
\midrule
64,000 & Random Testing & -34.35 & +0.0 & -- & -- & 1,250/25,000 & 0.050 & -- & -- \\
 & Random DisQ Testing & -37.78 & -10.0 & 0.00 & *** & 322/4,357 & 0.074 & 1.52 & *** \\
 & Genetic Testing & -38.35 & -11.6 & 0.26 & ** & 2,633/25,000 & 0.105 & 2.24 & *** \\
 & Genetic DisQ Testing & -44.15 & -28.5 & 0.00 & *** & 751/4,260 & 0.176 & 4.07 & *** \\
\midrule
128,000 & Random Testing & -30.86 & +0.0 & -- & -- & 1,250/25,000 & 0.050 & -- & -- \\
 & Random DisQ Testing & -34.91 & -13.1 & 0.00 & *** & 352/4,272 & 0.082 & 1.71 & *** \\
 & Genetic Testing & -38.28 & -24.0 & 0.07 & *** & 4,085/25,000 & 0.163 & 3.71 & *** \\
 & Genetic DisQ Testing & -39.79 & -28.9 & 0.00 & *** & 608/3,804 & 0.160 & 3.61 & *** \\
\bottomrule
\end{tabular}}
\end{table*}

\subsubsection{RQ3.4: Mid-episode Q-value Divergence}\label{sec:midepisode}

\paragraph{Setup}
DisQ evaluates the Q-value gap only at the initial state
(Section~\ref{sec:midepisode}), so it is blind to gaps that emerge later in
an episode, and more fundamentally to failures that never exhibit a
gap at all. This experiment measures both effects, contrasting an
environment where DisQ is effective (Simple Spread (2), RQ1) with the
environment where it is not (Knight Wizard Zombies, RQ3.2).

\paragraph{Execution}
For each environment and training regime, we executed $5{,}000$ episodes from uniformly sampled initial states under the deterministic trained policies and recorded $D(s_0)$ and $\max_t D(s_t)$, using the same threshold $\lambda$ calibrated from $D(s_0)$ as in RQ1.
We partitioned the failure episodes (5th-percentile criterion of Section~\ref{sec:metrics}) into three groups:
an initial-state gap ($D(s_0) > \lambda$, i.e., DisQ would select the episode),
a gap emerging only mid-episode ($D(s_0) \le \lambda$, $\max_t D(s_t) > \lambda$),
and no gap at any step.
As a reference for the predictiveness of mid-episode exceedance, we compute the rate at which $\max_t D(s_t) > \lambda$ among the non-failure episodes as well.

\paragraph{Result}
Table~\ref{tab:rq3_4_midepisode} reports the partition.
Since $\lambda$ is the $90\%$ quantile of $D(s_0)$, about $10\%$ of all episodes carry an initial-state gap by construction, so the fraction among failures measures whether the gap is informative: it is in Simple Spread (2) ($16.3\%$ and $22.3\%$, consistent with the enrichment in RQ1) but not in Knight Wizard Zombies ($10.7\%$ and $6.0\%$, at or below the base rate).
The remaining columns explain why. In Knight Wizard Zombies, most failures ($85.4\%$ centralized, $62.8\%$ decentralized) never exceed $\lambda$ at \emph{any} step: they are not accompanied by a Q-value gap, so no selection based on $D(s)$, initial or online, could target them, which explains the null result of RQ3.2 at the signal level.
In decentralized Simple Spread (2), by contrast, the gap accompanies most failures but is present at $s_0$ in only $22.3\%$, while $59.5\%$ exhibit it only mid-episode; initial-state selection thus captures an enriched but partial share, and online monitoring of $D(s_t)$ could recover more.
Whether such monitoring would be selective depends on the training regime: comparing how often $D(s_t)$ ever exceeds $\lambda$ in failure versus non-failure episodes, the decentralized policies show a $2.2$-fold elevation in both environments ($37.2\%$ vs.\ $16.9\%$; $81.8\%$ vs.\ $37.6\%$), while the centralized policies show none ($14.6\%$ vs.\ $15.5\%$; $26.2\%$ vs.\ $25.2\%$), consistent with their shared Q-network, for which $D(s)$ stems solely from differing observations passed through identical weights.
Online monitoring is therefore a promising extension for decentralized CMARL in future work.

\begin{table}[t]
  \centering
  \caption{Partition of failure episodes ($5{,}000$ uniformly sampled episodes per configuration) by when, if ever, the Q-value gap exceeds the DisQ threshold $\lambda$ ($90\%$ quantile of $D(s_0)$).\label{tab:rq3_4_midepisode}}
  \resizebox{\textwidth}{!}{%
  \begin{tabular}{llrrrr}
    \toprule
    Environment & Setting & Failures & $D(s_0) > \lambda$ & Mid-episode only & Never \\
    \midrule
    Knight Wizard Zombies & centralized   & 261 & 28 (10.7\%) & 10 (3.8\%)   & 223 (85.4\%) \\
    Knight Wizard Zombies & decentralized & 266 & 16 (6.0\%)  & 83 (31.2\%)  & 167 (62.8\%) \\
    Simple Spread (2)     & centralized   & 233 & 38 (16.3\%) & 23 (9.9\%)   & 172 (73.8\%) \\
    Simple Spread (2)     & decentralized & 291 & 65 (22.3\%) & 173 (59.5\%) & 53 (18.2\%) \\
    \bottomrule
  \end{tabular}}
\end{table}

\subsubsection{RQ3: Joint Answer}
DisQ's advantage is bounded along all four axes:
(1) It shrinks as the local reward ratio grows and vanishes in practical terms above a ratio of $0.9$ (RQ3.1);
(2) it disappears entirely, without degrading performance relative to the baseline, in environments where initial-state coordination has little influence on the episode outcome, with reward changes of at most $0.3\%$ in Knight Wizard Zombies, where the initial agent positions account for only $0.4\%$ of the reward variance (RQ3.2);
(3) it requires converged value estimates: for decentralized policies trained for fewer than $8{,}000$ steps, vanilla Genetic Testing detects lower rewards than its DisQ variant, and for intermediately trained centralized policies Random DisQ even selects fewer failures than random sampling ($\mathrm{OR}$ down to $0.53$), whereas for the fully trained policies DisQ again enriches failures significantly ($\mathrm{OR} = 1.58$, $p < 0.001$) (RQ3.3);
and (4) its selection signal is bounded by when and whether a gap occurs at all: in Knight Wizard Zombies, 63--85\% of failures never exhibit a gap, explaining the null result of RQ3.2 at the signal level, while in decentralized Simple Spread (2) the gap accompanies $82\%$ of failures but is present at $s_0$ in only $22\%$.
During execution, threshold exceedance is failure-predictive only for decentralized policies (RQ3.4).

\subsection{RQ4: What information does the Q-value gap provide for explaining selected test cases and observed failures?}

\paragraph{Setup}
This experiment examines what information the Q-value gap provides for explaining low-reward episodes.

\begin{figure*}[t]
    \centering
    \begin{subfigure}[b]{0.49\linewidth}
        \centering
        \includegraphics[width=\linewidth]{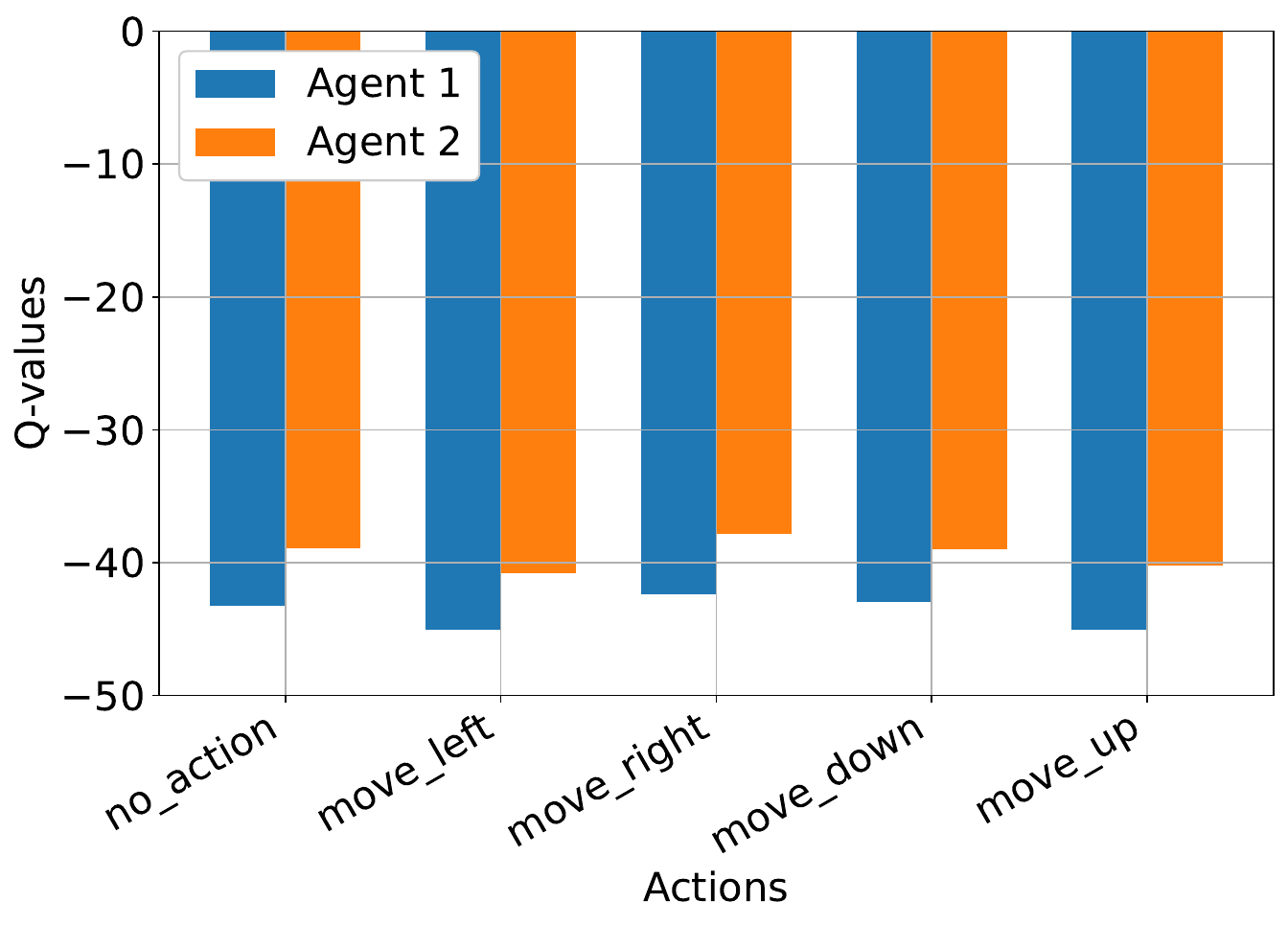}
        \caption{Centralized CMARL.}
        \label{fig:cen_q_values}
    \end{subfigure}
    \hfill
    \begin{subfigure}[b]{0.49\linewidth}
        \centering
        \includegraphics[width=\linewidth]{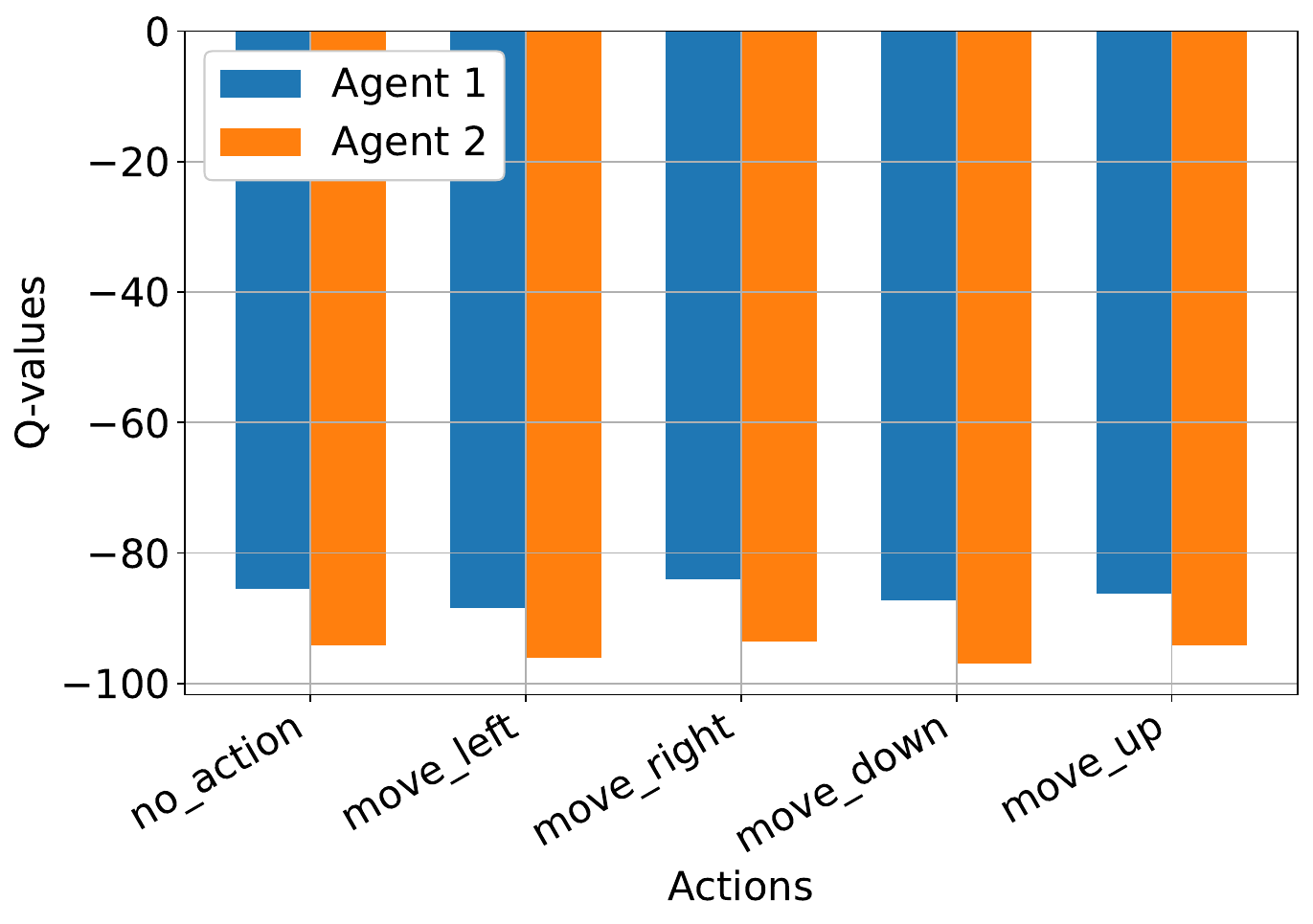}
        \caption{Decentralized CMARL.}
        \label{fig:dec_q_values}
    \end{subfigure}
    \caption{Q-values of agent 1 and agent 2 for each of their available actions in a single Q-value gap state of Simple Spread (2), selected by DisQ during testing. }
    \label{fig:q_values}
\end{figure*}

\paragraph{Execution}
In this experiment, we chose Q-value gap states from the previously executed tests and analyzed the Q-values of the trained CMARL policies.
We plotted the Q-values in Figure~\ref{fig:q_values}.

\paragraph{Result}
In Figure~\ref{fig:cen_q_values}, the centralized policy estimates an accumulated reward of around $-40$ for a given state at the end of the episode.
However, the maximum Q-values exhibit a difference of $4.5$, exceeding the empirically established threshold of $3.79$, indicating a Q-value gap between the agents.
After running the centralized policy from the initial state, we observed an episode reward of $-83.07$ (below the 5th-percentile threshold and therefore a failure), indicating a reward far from the Q-value predictions.

In Figure~\ref{fig:dec_q_values}, the decentralized policies estimate an accumulated reward of around $-95$ for a given state at the end of the episode.
However, the maximum Q-values exhibit a difference of $9.57$, exceeding the empirically established threshold of $6.79$, indicating a Q-value gap between the agents.
After running the decentralized policy from the initial state, we observed an episode reward of $-107$, indicating a failure.

In both of these cases, the Q-value gap indicated a failure of the agents and DisQ makes the selection straightforward to interpret, since it rests on expected rewards and a mismatch in those expectations among the agents.
Both the failure of the agents and the selection of the test case can be explained by this simple notion of the Q-value gap.

\paragraph{Answer}
Analyzing the Q-values and identifying Q-value gaps helps explain why an episode failed, namely that the agents held mismatched expectations about the return.
In the two inspected cases, gaps of $4.5$ and $9.57$ (against thresholds of $3.79$ and $6.79$) preceded episode rewards of $-83.07$ and $-107$, far below the agents' own predictions of roughly $-40$ and $-95$.
Looking at the Q-value gap provides a simple and interpretable way to explain the test case selection itself, since it is only a single number representing said mismatch between agents, without requiring further instrumentation.

\section{Threats to Validity}\label{sec:threats}
A first threat to validity is that the study focuses on certain environments (i.e., Multi-Agent Particle environments, Coin Collector, and the Knight Wizard Zombies game).
While we select a variety of different environments, these environments might not fully represent the diversity of all CMARL applications.

A second threat is that implementation bugs of the DisQ method could introduce errors in the results, even though the DisQ logic is straightforward.

A third threat concerns the operationalization of failures.
Our discrete failure metric instantiates Definition~\ref{def:failure} with a fixed reward threshold, the 5th percentile of the baseline reward distribution per policy under test.
Unsafe events enter this criterion only through their reward penalties: failures whose safety impact is not (or is only weakly) penalized by the reward function would not be counted, which is a general limitation of reward-based oracles rather than of DisQ specifically.
A different percentile would change the absolute failure counts, but the reported odds ratios measure the \emph{enrichment} of failure episodes relative to the same threshold applied to the baseline and are therefore comparable across methods and environments.

The more consequential limitation is which \emph{class} of failure the metric captures: as noted after Definition~\ref{def:coordfailure}, a reward threshold cannot separate a coordination failure from one in which a single agent acts poorly, so our claims are about reward-threshold failures rather than coordination failures specifically.
The two are nonetheless closely associated empirically.
We compared the reward-based label against threshold-free coordination events in two environments: a collision between agents in Simple Spread (2), and the listener ending nearest a landmark other than the speaker's goal in Speaker Listener.
Over $25{,}000$ uniformly sampled episodes per configuration, collisions occur in $22.1\%$ of failure episodes against $15.9\%$ of non-failure episodes under centralized training, and in $47.6\%$ against $21.7\%$ under decentralized training.
The corresponding rates for the wrong landmark are $98.6\%$ against $64.5\%$ and $99.2\%$ against $64.4\%$.
The wrong-landmark event is frequent even among non-failure episodes, so its base rate is high and its enrichment correspondingly modest.
Reward-threshold failures are thus consistently enriched in observable coordination events, by factors of $1.4$ to $2.2$ across all four comparisons, without the label isolating them.
Beyond this, the metric does not distinguish qualitatively different failure types, so we cannot assess the diversity of the exposed failures.

A fourth threat concerns the genetic baseline rather than DisQ.
Its individuals encode only the initial agent positions, while the rest of the initial state is re-drawn on every reset, so each fitness value is a single-sample estimate.
With those positions held fixed and the remainder re-drawn, they account for $7.0\%$ of the episode-reward variance in Simple Spread (2), $3$--$4\%$ in Coin Collector and $0.4\%$ in Knight Wizard Zombies; where the share is small the search is driven largely by sampling noise, which is the most plausible explanation for the genetic results in Coin Collector (Section~\ref{sec:results}).
Encoding the remaining elements of the initial state would give the search more to act on, so the reported genetic results are a lower bound on what genetic SBST could achieve in these environments.

Finally, the performance of all the testing methods may also be sensitive to the specific hyperparameters used for training the agents, which were manually tuned in this study.
A different set of hyperparameters might yield different outcomes, potentially leading to different conclusions about the method's~effectiveness.
Because DisQ operates directly on learned Q-values, its behavior depends on training choices that shape those values.
The discount factor $\gamma$ compresses or stretches the scale of Q-values and therefore of $D(s)$.
However, since $\lambda$ is calibrated as an empirical quantile of $D(s)$ for the specific policies under test, a global rescaling of Q-values leaves the selection invariant.
Only the interpretability of the absolute gap magnitude changes.
Non-uniform effects (e.g., $\gamma$ interacting differently with each agent's horizon, or differing reward normalizations per agent) are not covered by this argument and could affect DisQ.
RQ3.3 covers one such axis (training duration), while a systematic sweep over training hyperparameters remains future work.
To mitigate threats to conclusion validity, all comparisons are performed on independent repetitions rather than pooled episodes, paired with effect sizes, and corrected for multiple comparisons (Section~\ref{sec:metrics}).

\section{Related Work}\label{sec:relwork}

\subsection{Explainable CMARL}
Boggess et al. summarize policies to explain agent cooperation by abstracting the environment and policies, and provide language explanations to answer queries about agent behavior such as ``When do [agents] do [actions]?'', ``Why don't [agents] do [actions] in [states]?'' and ``What do [agents] do in [predicates]?'' \cite{DBLP:conf/ijcai/BoggessK022}.
In contrast, our method identifies potential miscoordination by analyzing the Q-value gap for specific states.
Heuillet et al. present global model-agnostic quantitative explanations about the individual contribution of agents in a cooperative multi-agent RL environment \cite{DBLP:journals/cim/HeuilletCR22}.
Li et al.~\cite{DBLP:conf/icml/LiLC024} and Chen and Tan~\cite{DBLP:journals/nn/ChenT23} follow a similar approach.
In contrast, our method offers a localized, fine-grained perspective by pinpointing state-specific inconsistencies in agent coordination rather than providing an overarching quantitative assessment.

Boggess et al.~\cite{DBLP:conf/ijcai/BoggessK023} present an approach for generating policy-level contrastive explanations for MARL to answer a temporal user query, which specifies a sequence of tasks completed by agents with possible cooperation.
The proposed approach encodes the temporal query as a PCTL logic formula and checks if the query is feasible under a given MARL policy via probabilistic model checking.
Gross et al.~\cite{DBLP:conf/aips/GrossS0023} take a similar approach.
In contrast, our method does not rely on formal logic encoding.

Orner et al. present a method called SICA \cite{DBLP:journals/corr/abs-2407-15255}.
SICA helps explain how and why agents might cooperate or compete in a given situation by analyzing over multiple simulations the correlation of their reward outcomes when they follow their own policies.
It provides a high-level view of the relationships between agents, offering insights into their mutual ``friendliness'' or ``animosity'', which is valuable for understanding and designing multi-agent systems.
Instead of evaluating long-term reward correlations, our method detects localized inconsistencies in agent coordination at the decision-making level.

Ho and Wang present an explainable and adaptable augmented knowledge attention network architecture for multi-agent DRL systems \cite{DBLP:conf/aike/HoW20}.
In contrast to our work, they investigate how the policies interpret the inputs, while we focus on investigating the output of the neural network (NN) policies.

A well-known approach is action advising, where a teacher provides advice as state-action pairs to guide the student. While successful in simple environments, action advising often struggles in complex environments with large state spaces.
To address this, Guo et al. propose an algorithm where the teacher provides both action advice and a path explanation, showing that incorporating these explanations of an internal decision tree policy enables learning through reflection \cite{DBLP:conf/icra/GuoCSLHFS23}.
In comparison, we use our explainable CMARL metric for testing policies instead of training them.

Trajectory prediction is the task of predicting an agent's future states, which can be seen as a kind of explanation.
Wu et al.~\cite{DBLP:conf/corl/WuCGBM23} propose the DTDE MARL algorithm for joint trajectory and intent prediction for autonomous vehicles in dense and heterogeneous environments.
Their algorithm is fully decentralized without weight sharing, communication, or centralized critics, and can handle variable agents across episodes.
While they predict intents, we predict failures.

Lazy agent pathology occurs when agents receive incorrect credit assignments~\cite{DBLP:conf/atal/SunehagLGCZJLSL18}, leading them to learn suboptimal policies and not cooperate towards the team's common goal.
Detecting lazy agents can be viewed as a form of explainability.
Pina et al. introduced Independent Causal Learning (ICL) and demonstrated that detecting causal relations between individual observations and rewards in MARL can detect and address this problem \cite{DBLP:conf/atal/PinaSA23}.
While lazy agents can be seen as an explainable CMARL approach in the context of failures, we focus on the discrepancy of future predictions of the individual policies to each other.

\subsection{Single- and Multi-agent RL Testing}
Single-agent RL policy testing, the closest neighbor of CMARL testing, has received extensive attention.
\emph{DeepTest}~\cite{DBLP:conf/icse/TianPJR18} is an automated testing framework for NNs in autonomous driving, utilizing metamorphic testing.
It specifically targets models that process image inputs. Similarly, DeepXplore tests image-based NN agents but uses differential testing to identify errors~\cite{10.1145/3361566}. In DeepXplore, multiple policies are provided, and failures are detected by comparing their outputs.

\emph{MDPFuzz} is a black box fuzz testing framework for policies tackling MDPs, testing if the target policy reaches abnormal or dangerous states~\cite{DBLP:conf/issta/PangYW22,10.1145/3650212.3680382}.
It features a cover model of the state sequences observed during testing to guide the fuzzing process toward unseen behaviors.
Similarly to \emph{MDPFuzz}, Eniser et al.~\cite{DBLP:conf/issta/EniserGW0C22} and Eisenhut et al.~\cite{Eisenhut2023AutomaticMT} employ fuzzing techniques to generate test inputs, but the results are subsequently tested using metamorphic testing.
Recently, Mazouni et al.~\cite{DBLP:conf/ast/MazouniSGA24} and Bhatt et al.~\cite{bhatt2022deepsurrogateassistedgeneration} explore the use of Quality-Diversity optimization~\cite{10.3389/frobt.2016.00040} to search for failures.

On the other hand, RL itself has been used to optimize the search.
For instance, DeepCollision trains an RL agent to modify the environment during the execution of the test cases to detect failures~\cite{9712397}.
Likewise, Haq et al.~\cite{10.1109/ICSE48619.2023.00155} propose MORLOT, a framework for handling many-objective policy testing tasks.
Their approach consists of training a deep RL agent~\cite{mnih2013playing} to find failure-triggering configurations of the environment.

Other types of SBST approaches include genetic algorithm~\cite{zolf23}, depth-first search~\cite{tappler2022searchbasedtestingreinforcementlearning}, and test oracle approximation through AI planning heuristics~\cite{DBLP:conf/aips/SteinmetzFEFGHH22}.
We refer the interested reader to the survey by Mazouni et al.~\cite{DBLP:journals/corr/abs-2312-09680} for a broader review of policy testing methods.

While we do not focus on adversarial testing, adversarial testing has also been extensively applied to RL policy testing.
The first line of research aims to reduce the number of times adversaries are injected (while achieving similar results to systemic attacks).
For instance, Kos and Song~\cite{kos2017delvingadversarialattacksdeep} refine FGSM (Fast Gradient Sign Method) by only attacking critical states during the episodes, which they select based on the value function of the policy.
On the other hand, Lin et al.~\cite{lin2019tacticsadversarialattackdeep} identify when to attack the policy by targeting states that maximize the difference between the most preferred action and the least preferred action.
Sun et al.~\cite{sun2020stealthyefficientadversarialattacks} opt for another strategy for planning the attacks and emphasize long-term damage impacts.
They propose to achieve this goal by learning an antagonist model, which they train with the reward function to account for the agent's end goal.
Another line of work aims at the adversaries themselves.
For instance, Chan et al.~\cite{10.1145/3320269.3384715} quantify how much changing each feature composing the observations affects the rewards with a \emph{static reward impact map} to only perturb the most impactful features.
Similar to testing, prior work also formally verifies adversarial attacks on RL policies in both single-agent~\cite{DBLP:conf/icaart/GrossS0023} and multi-agent settings~\cite{DBLP:conf/aips/GrossS0023}.

\emph{COOL-MC} enables training single-agent and multi-agent RL policies in modeled environments while formally verifying them at any stage of the training process~\cite{DBLP:conf/setta/GrossJJP22}.
\emph{MoGym} is an integrated toolbox that trains and verifies RL agents using deep statistical model checking~\cite{DBLP:conf/tacas/KohlKH21}.
\emph{Mungojerrie} is a tool used for testing reward functions, specifically for $\omega$-regular objectives~\cite{DBLP:journals/corr/abs-2106-09161}.
Gu et al. introduce MCRL, merging verification with MARL for mission planning to ensure safety-critical requirements~\cite{DBLP:journals/sttt/GuJPSEL22}.
The main difference is that we focus on testing instead of verifying CMARL policies.
\emph{Gimitest} is a tool for testing single- and multi-agent RL policies in Farama Gymnasium, PettingZoo, and similar framework environments.
It streamlines testing by extending the environment simulators and supports customizable test cases and configurations~\cite{gimitest}.
We use Gimitest to implement our testing and test case selection methods.

Zhou et al. present a MARL testing method that tests if the trained agents are robust against noise state observations~\cite{DBLP:conf/ecai/ZhouL23} by presenting a method called RTCA.
RTCA introduces two key innovations: 1) a Differential Evolution (DE) method to select critical agents as noise victims and determine worst-case joint actions; 2) a team cooperation policy evaluation used as the DE optimization objective. Adversarial state perturbations are then applied to the critical agents based on these joint actions.
In contrast, our method focuses on identifying miscoordination by analyzing Q-value gaps in specific and clean states.
Ma and Li~\cite{DBLP:conf/atal/MaL23}, Liu and Lai~\cite{DBLP:conf/nips/LiuL23}, Figura et al.~\cite{DBLP:conf/amcc/FiguraK021}, and Lv et al.~\cite{DBLP:journals/tifs/LvXCCJ24} study adversarial testing in multi-agent settings.
The main difference is, as previously mentioned, that we focus on safety instead of security violations.

Recently, Ma et al.~\cite{DBLP:conf/issta/MaWWXWLX024} introduced a multi-agent system testing method, including multi-agent RL, that focuses on finding diverse test cases. To this end, multiple diversity measures are combined, e.g., individual and team diversity of the agents, failure degrees, and frequencies, to build a diverse set of test cases along the several diversity dimensions.
New test cases are generated and identified through action perturbations and the abstract representation of states and trajectories before the diversity measures are calculated.
While shown to be very effective in several case studies, the method is much more involved and complex than our proposed DisQ method.
For example, DisQ's selection only relies on the initial state and the agents' corresponding Q-values, but does not require the actual execution of the episode from this state including a post-episode analysis of the trajectories.
These two methods are, therefore, substantially different in their approach and focus.

\subsection{Combining Testing and Explainability}
The final, closely related category is a combination of testing and explainable RL methods.
Biagiola et al.~\cite{DBLP:journals/tosem/BiagiolaT24} propose an SBST approach to test agents. Their approach, implemented in a tool called Indago, trains a classifier on failure and non-failure environment (i.e., pass) configurations resulting from the DRL training process. 
The classifier is used during testing as a surrogate model for DRL agent execution in the environment, predicting the likelihood that a given environment configuration will induce a failure in the DRL agent under test. 
This failure prediction serves as a fitness function, guiding the generation towards failure-prone configurations while saving computation time by limiting DRL agent execution to those configurations most likely to expose failures. 
They use explainable methods to steer the evolutionary testing algorithm.
On the other hand, we focus on CMARL and consider the dynamics of the CMARL policies.

In the context of single-agent explainable and verified RL, existing work iteratively prunes NN policies to interpret the feature importance for safety at a global level~\cite{SafetyPruningESANN2024}.
Some work leverages large language models to identify safety-critical states and apply counterfactual reasoning to explain why the RL policy violated a safety property while also proposing alternative actions~\cite{gross2024enhancing}.
Existing work uses model checking to query multi-agent RL agent behaviors and uses this extracted behavior \emph{explicitly} as an explanation~\cite{DBLP:conf/ijcai/BoggessK023}.

\section{Conclusion}\label{sec:conclusion}
This paper introduces a test case selection method called DisQ (Disagreeing Q-values), which aims to improve failure detection in parallel CMARL systems.
The method identifies Q-value gaps: states in which the agents' maximal Q-values differ by more than a calibrated threshold.
These gaps indicate that the agents are not aligned in their decision-making, suggesting potential miscoordination or failure behavior throughout the episode, and we identify the conditions under which the signal is absent.

The main benefit of DisQ is that it is straightforward to apply because it relies on Q-values for an initial state, which are already calculated in many CMARL systems.
It also integrates with most existing test case generators, such as random search or genetic SBST, without major changes to the testing process.
By focusing on likely failure cases, it reduces unnecessary test executions and improves testing efficiency.

For future work, we plan to investigate the application of DisQ to policy gradient methods, which do not rely on Q-values but action probabilities~\cite{DBLP:journals/corr/SchulmanWDRK17}, perhaps by measuring the divergence between agents' action probability distributions, or, for MADDPG-style architectures, between the agents' critics evaluated at the joint greedy action.
We will consider how DisQ can be applied to turn-based cooperative MARL systems~\cite{DBLP:conf/icaart/Gross23} or how to make use of online monitoring of the Q-value gap.
To assess the practical utility of DisQ, we aim to apply it to real-world, industrial CMARL settings, such as autonomous vehicle coordination, distributed robotics, and collaborative scheduling systems.
Finally, we see an opportunity to investigate alternative explainable CMARL methods to improve the testing method efficiency.

\section*{Data Availability Statement}

The source code for the experiments of this study is openly available on GitHub at \url{https://github.com/DennisGross/cmarl_testing}.

\section*{Acknowledgments}
This work is funded by the European Union under grant agreement number 101091783 (MARS Project) and the Norwegian Ministry of Education and Research.

\bibliographystyle{unsrt}
\bibliography{refs}

\end{document}